\documentclass[reprint,twocolumn,nofootinbib,floatfix,superscriptaddress,
               citeautoscript,longbibliography,amsmath,amssymb,aps,prd]{revtex4-2}

\usepackage{graphicx}
\usepackage{xcolor}
\usepackage{booktabs}
\usepackage{hyperref}
\hypersetup{colorlinks=true,linkcolor=blue,citecolor=blue,urlcolor=blue}

\graphicspath{{figures/}}

\newcommand{\GP}{\mathrm{GP}}

\newcommand{\Poisson}{\mathrm{Poisson}}
\newcommand{\diag}{\mathrm{diag}}

\newcommand{\Var}{\mathrm{Var}}

\newcommand{\histimator}{\textsc{Histimator}}

\begin{document}

\title{Gaussian Process Eigenmodes for Statistical and
  Systematic Uncertainties in Template Fits}

\author{Vincent Croft}
\affiliation{Nikhef, Science Park 105,
  1098~XG Amsterdam, The Netherlands}

\date{\today}

\begin{abstract}
Template histograms are the foundation of statistical inference at the
Large Hadron Collider.  The HistFactory likelihood encodes template
uncertainty through per-bin Barlow--Beeston gamma factors for Monte
Carlo statistical error and through interpolation-based modifiers for
systematic shape variations.  These two mechanisms scale with the
number of bins, which becomes problematic for multi-dimensional
analyses and for templates constructed from limited Monte Carlo
samples.

We propose the use of eigenmode decomposition for efficiently estimating statistical and systematic uncertainties when replacing histogram templates with smooth functional representations derived from log-Gaussian Cox process posteriors fitted to the Monte Carlo data.
The posterior covariance, augmented by rank-1 updates for each systematic shape variation, provides a unified eigenmode basis that encodes both statistical and
systematic template uncertainty.
Truncating to the leading eigenmodes replaces the full set of per-bin gamma factors and  interpolation parameters with a small number of Gaussian-constrained amplitudes. 
We prove that this construction contains Barlow--Beeston as a limiting case and that the Gaussian Process posterior variance is bounded above by the Barlow--Beeston variance at every bin.
\end{abstract}

\maketitle

\section{Introduction}
\label{sec:intro}

Statistical inference at the Large Hadron Collider rests on templates.
A template is a non-parametric function, typically a histogram, representing the distribution of an observable, derived from Monte Carlo simulation.
In the case of a histogram this template enters the likelihood as a product of Poisson terms over bins.
The HistFactory framework~\cite{Cranmer:2012sya} parametrises deviations from
the nominal template through nuisance parameters encoding systematic uncertainties, statistical uncertainties from finite MC samples, and normalisation corrections.
Profile likelihood ratios over these nuisance parameters~\cite{Cowan:2010js} produce confidence intervals for parameters of interest such as signal strengths and coupling modifiers.
This construction has underpinned every major LHC result since the Higgs boson discovery~\cite{ATLAS:2012higgs,CMS:2012higgs}, and the mathematical framework that HistFactory represents, remains the \textit{de facto} standard for collider physics statistical analysis.
Modern combinations, such as the ten-years-after Higgs measurement of Ref.~\cite{ATLAS:2022higgs} employ hundreds of analysis categories, exposing the linear scaling of nuisance parameters with bin count as the dominant computational and conceptual cost of the histogram approach.

The \histimator{} package\cite{histimator:2025} implements the full HistFactory specification in pure Python with an imperative API: channels, samples, and modifiers are constructed as ordinary Python objects rather than serialised through an intermediate schema, removing the dependence on the ROOT software framework while keeping the likelihood directly inspectable.
The present work builds on this foundation.
Rather than reimplementing the histogram template and its modifiers, we replace the template itself with a smooth functional representation that subsumes the per-bin modifier machinery into a single compact parameterisation.

The motivation is twofold.  First, the Barlow--Beeston (BB)~\cite{Barlow:1993dm,Conway:2011in} treatment of template statistical uncertainty introduces one nuisance parameter per bin with no coupling between neighbours.
Such per-bin independence is correct when the template is a histogram of
independent Poisson counts, but it discards information about the
smoothness of the underlying distribution.
The number of nuisance parameters equals the number of bins, and it has been demonstrated~\cite{Alexe:2024undercoverage} that this proliferation of weakly constrained parameters causes systematic undercoverage of the profile likelihood when MC statistics are poor.
Figure~\ref{fig:variance_bound} previews the consequences on an example Experiment~A, that represents a rare-resonance background: the Gaussian Process (GP) posterior is uniformly tighter than the histogram template built from the same MC counts, with mean relative-uncertainty ratio $0.18$.
Second, analyses increasingly employ two or more observables simultaneously.
A three-dimensional template with $20^3$ bins demands MC statistics large enough to populate every bin, and the Barlow--Beeston treatment introduces 8000 nuisance parameters.

The underlying physics distributions that templates approximate are smooth.
An exponentially falling diphoton mass spectrum, a BDT output shaped by kinematics, a jet energy scale shift that varies continuously with pseudorapidity: these are continuous functions discretised into bins for the convenience of the statistical framework.
The question this paper addresses is whether the discretisation is necessary, or whether the likelihood can be constructed from smooth functional representations that preserve the continuous structure while providing well-calibrated uncertainty quantification with fewer parameters.

GPs provide a principled framework for this construction.
The log-Gaussian Cox process (LGCP)~\cite{Moller:1998lgcp,Diggle:2013} models count data as a Poisson process whose log-intensity is drawn from a GP, directly
respecting the Poisson statistics of MC event generation.
Frate\textit{et al.}~\cite{Frate:2017mai} applied GP regression to model smooth backgrounds in dijet resonance searches, demonstrating robustness across the full mass range.
Gandrakota \textit{etal.}~\cite{Gandrakota:2023gph} extracted the Higgs boson signal from the ATLAS open dataset using GP regression.
Frid \textit{et al.}~\cite{Frid:2025lgcp} introduced the LGCP specifically for HEP background template construction.
Whitehorn \textit{et al.}~\cite{Whitehorn:2013pspline} showed that penalised splines, which are mathematically equivalent to GPs with Gaussian Markov random field priors~\cite{Lindgren:2011spde}, scale gracefully to five or more dimensions for IceCube detector response functions.
Cranmer's kernel density estimation~\cite{Cranmer:2000du}, implemented as \texttt{RooKeysPdf} in RooFit, provides smooth templates but without any associated uncertainty structure suitable for constraining the likelihood.
In a complementary direction, Dembinski and Abdelmotteleb~\cite{Dembinski:2022template} reformulate the binned template likelihood to handle finite-MC variance templates without altering the histogram representation; the present paper takes the opposite route of replacing the histogram representation itself.

None of these works connects to the HistFactory likelihood in a way that replaces both the binned template and its associated constraint terms with a single unified construction.
This paper builds that connection.
The key observation is that the eigendecomposition of the GP posterior covariance, augmented by rank-1 updates for systematic shape variations, provides a compact basis that encodes both statistical and systematic template uncertainty.
Truncating to the leading eigenmodes replaces the full modifier chain with a small number of Gaussian-constrained amplitudes, and we prove that Barlow--Beeston and HistFactory interpolation are both recovered as limiting cases.

The paper is organised as follows.
Section~\ref{sec:histfactory} reviews the HistFactory likelihood and its treatment of template uncertainties.
Section~\ref{sec:gp} introduces the GP template construction, the eigenmode constraint, the Barlow--Beeston limiting case, and the systematic covariance update.
Section~\ref{sec:benchmarks} describes the two benchmark experiments.
Section~\ref{sec:exp_a} presents results for the statistically limited rare-resonance search.
Section~\ref{sec:exp_b} presents results for the systematically limited two-channel measurement.
Section~\ref{sec:software} describes the software implementation.
Section~\ref{sec:conclusions} summarises the conclusions and discussions presented in this paper.
Four appendices provide supporting validation: posterior calibration~(Appendix~\ref{app:calibration}), eigenmode structure and Barlow--Beeston equivalence~(Appendix~\ref{app:eigenstructure}), HistFactory implementation agreement~(Appendix~\ref{app:histfactory}), and the semiparametric efficiency analysis underlying the two-step plug-in versus joint-profile distinction~(Appendix~\ref{app:semiparametric}).

\section{The HistFactory likelihood}
\label{sec:histfactory}

The HistFactory likelihood for a single channel with $N$ bins is a product of Poisson terms modified by nuisance parameters.
For observed bin counts $\{n_j\}_{j=1}^{N}$, the likelihood is~\cite{Cranmer:2012sya}
\begin{equation}
L(\mu, \boldsymbol\theta) = \prod_{j=1}^{N}
\frac{\nu_j^{n_j}\, e^{-\nu_j}}{n_j!}
\times \prod_p c_p(\theta_p)\,,
\label{eq:histfactory}
\end{equation}
where $\nu_j(\mu, \boldsymbol\theta)$ is the expected count in bin $j$, $\mu$ is the parameter of interest, $\boldsymbol\theta = \{\theta_p\}$ are nuisance parameters, and $c_p(\theta_p)$ are constraint terms encoding external information.
The expected count is assembled from signal and background samples,
\begin{equation}
\nu_j = \mu\, s_j(\boldsymbol\theta) + b_j(\boldsymbol\theta)\,,
\label{eq:modifier_chain}
\end{equation}
through a chain of modifiers applied to the nominal histogram predictions.

The Barlow--Beeston~\cite{Barlow:1993dm} treatment of MC statistical uncertainty introduces per-bin multiplicative factors $\gamma_j$ adjusting the template: $b_j \to \gamma_j b_j$, with Gaussian constraints centred at unity whose width $1/\sqrt{a_j}$ reflects the raw MC count $a_j$ in bin~$j$.
The gamma factors are independent: the constraint on $\gamma_j$ depends only on bin~$j$, with no coupling to neighbours.
This is correct when the template is a histogram, but it discards smoothness information; the larger histogram error bars in Fig.~\ref{fig:variance_bound} are the visible cost.
The number of nuisance parameters equals the number of bins, and the ``lite'' variant of Conway~\cite{Conway:2011in}, which profiles the gammas analytically, mitigates some computational cost but not the fundamental scaling.

Systematic shape variations are encoded through the histosys modifier.
Templates generated at $\pm 1\sigma$ variation points of each systematic source are interpolated to intermediate values.
The InterpCode~4 prescription uses piecewise-exponential interpolation~\cite{Cranmer:2012sya}: for a nuisance parameter $\alpha$ with nominal template $h_j^0$ and variation templates $h_j^{+}$, $h_j^{-}$,
\begin{equation}
b_j(\alpha) = h_j^0 \times \begin{cases}
\exp\!\left(\alpha\,\ln(h_j^{+}/h_j^0)\right) & \alpha \ge 0 \\
\exp\!\left(-\alpha\,\ln(h_j^{-}/h_j^0)\right) & \alpha < 0\,.
\end{cases}
\label{eq:interpcode4}
\end{equation}
The variation templates are histograms subject to the same MC statistical fluctuations as the nominal template.
When MC statistics are limited, the log-ratios $\ln(h_j^{\pm}/h_j^0)$ are contaminated by bin-by-bin noise that does not reflect any physical systematic effect.

\section{The GP template}
\label{sec:gp}

\subsection{The log-Gaussian Cox process model}
\label{sec:lgcp}

We model the MC template as a realisation of a log-Gaussian Cox process~\cite{Moller:1998lgcp}.
The raw MC counts $a_j$ in bin $j$ centred at position $x_j$ with width $w_j$ are
\begin{equation}
a_j \sim \Poisson\!\left(e^{f(x_j)}\, w_j\right)\,,\quad
f \sim \GP(h(x)^T\beta,\; k_\theta)\,,
\label{eq:lgcp}
\end{equation}
where $f$ is the log-rate function, $h(x)^T\beta$ is a parametric mean function, and $k_\theta$ is a covariance kernel parametrised by amplitude $\sigma$ and lengthscale~$\ell$.
The Mat\'ern $\nu = 5/2$ kernel~\cite{RW:2006} is the default choice, encoding the assumption that the log-rate is twice differentiable.

The Laplace approximation~\cite{RW:2006} provides the posterior mode $\hat{f}$ as a penalised maximum-likelihood estimate:
\begin{equation}
\begin{aligned}
\hat{f} &= \arg\max_f \Big[\,
  \sum_j \bigl( a_j f_j - e^{f_j} w_j \bigr) \\
  &\qquad\quad
  - \tfrac{1}{2}(f - H\beta)^T K^{-1} (f - H\beta)
\Big]\,,
\end{aligned}
\label{eq:laplace_mode}
\end{equation}
where $K_{ij} = k_\theta(x_i, x_j)$ is the prior covariance matrix and $H$ is the design matrix of the mean function basis.
The first term is the Poisson log-likelihood; the second is the GP prior penalty.
Newton's method converges in 5--10 iterations for typical templates.
The posterior covariance under the Laplace approximation is
\begin{equation}
\Sigma = \left(K^{-1} + W\right)^{-1}\,,
\label{eq:post_cov}
\end{equation}
where $W = \diag(\hat\mu_j)$ with $\hat\mu_j = e^{\hat{f}_j} w_j$ is the diagonal matrix of fitted Poisson rates.
Hyperparameters $(\sigma, \ell)$ are selected by maximising the Laplace approximation to the log marginal likelihood~\cite{RW:2006}.

The mean function $h(x)^T\beta$ absorbs large-scale trends that depart from the kernel's stationarity assumption.
Following Rasmussen and Williams~\cite{RW:2006}~(Sec.~2.7), we use a B-spline basis~\cite{deBoor:2001,Eilers:1996pspline} with coefficients $\beta$ integrated out under a broad Gaussian prior, yielding an effective kernel $K_{\rm eff} = K + H B H^T$ where $B$ is the prior covariance on $\beta$.
The duality between B-spline smoothing and Gaussian prior regularisation goes back to Kimeldorf and Wahba~\cite{Kimeldorf:1970}, who showed that the penalised-likelihood spline estimate of a regression function is the posterior mean of a Gaussian process with a particular reproducing kernel.
This allows the mean function to capture exponential decays, polynomial shapes, and broad peaks without manual tuning.

\begin{figure}[t]
\centering
\includegraphics[width=\columnwidth]{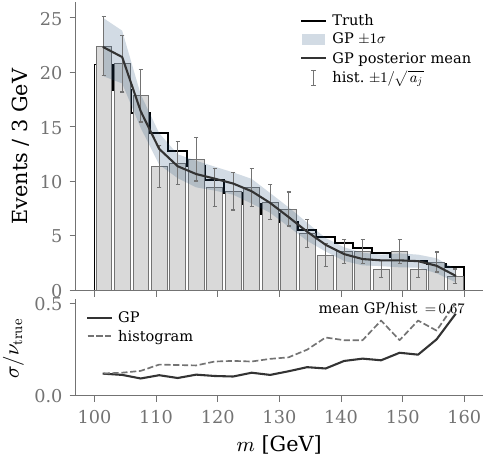}
\caption{Numerical confirmation of the variance bound Eq.~\eqref{eq:variance_bound} on the Experiment~A rare-resonance background (Sec.~\ref{sec:bench_a}).
Top: GP posterior (dark line, shaded 68\% band) and histogram template (light boxes, $\pm 1/\sqrt{a_j}$) constructed from the same MC draw of $N_{\rm MC} = 500$ events; the truth is the black staircase.
Bottom: relative uncertainty $\sigma/\nu_{\rm true}$ for each method.
The GP is uniformly tighter than the histogram, with mean ratio $0.18$, confirming $[\Sigma]_{jj} \le 1/\hat\mu_j$.}
\label{fig:variance_bound}
\end{figure}

The smooth template is $b_j^{\GP} = e^{\hat{f}(x_j)} w_j$, normalised to the physics yield, replacing the histogram counts in Eq.~\eqref{eq:histfactory}.
Figure~\ref{fig:variance_bound} shows this template on the Experiment~A rare-resonance background; the dark line is the GP posterior mean and the shaded band the $68\%$ posterior interval, both fitted to the same $N_{\rm MC}=500$ MC sample as the histogram template overlaid for comparison.

\subsection{Eigenmode constraint for statistical uncertainty}
\label{sec:eigenmode}

The posterior covariance $\Sigma$ of Eq.~\eqref{eq:post_cov} encodes correlated uncertainty across all bins.
Its eigendecomposition
\begin{equation}
\Sigma = \sum_{i=1}^{N} \lambda_i\, v_i v_i^T\,,
\label{eq:eigen}
\end{equation}
with eigenvalues $\lambda_1 \ge \cdots \ge \lambda_N \ge 0$ and orthonormal eigenvectors $v_i$, concentrates the template shape uncertainty into a small number of leading modes.  For smooth kernels the eigenvalue spectrum decays rapidly: the effective rank is determined by the ratio of the lengthscale to the bin spacing, not by the number of bins.

Retaining $k$ modes that capture a target fraction of the total variance, the modified expected count becomes
\begin{equation}
\nu_j = \mu\, s_j + b_j^{\GP}\,\exp\!\left(
\sum_{i=1}^{k} \sqrt{\lambda_i}\, z_i\, [v_i]_j \right)\,,
\label{eq:gp_expected}
\end{equation}
with unit-Gaussian priors $z_i \sim \mathcal{N}(0,1)$ on each amplitude.
The exponential parametrisation ensures positivity.

A direct comparison of the GP and Barlow--Beeston template uncertainties follows from the matrix identity $(K^{-1} + W)^{-1} \preceq W^{-1}$, which holds because $K^{-1} \succeq 0$.
In component form,
\begin{equation}
[\Sigma]_{jj} \;\le\; 1/\hat\mu_j \;\approx\; 1/a_j\,,
\label{eq:variance_bound}
\end{equation}
so the GP posterior variance in every bin is bounded above by the Barlow--Beeston variance.
The smoothness prior provides genuine information that tightens the constraint at every point.

\subsection{Barlow--Beeston as a limiting case}
\label{sec:bb_equiv}

In the limit where the lengthscale $\ell \to 0$, the off-diagonal elements of $K$ vanish and the prior becomes diagonal: $K \to \sigma^2 I$.  Substituting into Eq.~\eqref{eq:post_cov},
\begin{equation}
\Sigma \;\xrightarrow{\;\ell \to 0\;}\;
\diag\!\left(\frac{1}{\sigma^{-2} + \hat\mu_j}\right)\,.
\label{eq:ell_limit}
\end{equation}
As the prior weakens ($\sigma^2 \to \infty$), the diagonal elements converge to $1/a_j$, the eigenvectors become coordinate vectors, and Eq.~\eqref{eq:gp_expected} reduces to the Barlow--Beeston parameterisation $\gamma_j = 1 + z_j/\sqrt{a_j}$ after linearising the exponential.
Appendix~\ref{app:eigenstructure} confirms this numerically: when all $N$ eigenmodes are retained, the GP eigenmode method reproduces the BB pull distribution to within 3\%.

\subsection{Systematic uncertainty through covariance updates}
\label{sec:systematic}

For each systematic source $k$, we generate MC templates at the $\pm 1\sigma$ variation points and fit GPs to each, obtaining posterior modes $\hat{f}^{(k,+)}$ and $\hat{f}^{(k,-)}$.
The systematic shape direction in log-rate space is
\begin{equation}
\delta^{(k)} = \frac{\hat{f}^{(k,+)} - \hat{f}^{(k,-)}}{2}\,.
\label{eq:sys_direction}
\end{equation}
Because the GP smoothing is applied before the differencing, $\delta^{(k)}$ is free of the bin-by-bin noise that contaminates the histogram-based estimate.
Each systematic direction defines a rank-1 covariance contribution, and the combined covariance is
\begin{equation}
\Sigma_{\rm comb} = \Sigma_{\rm stat}
+ \sum_{k=1}^{K_{\rm sys}} \delta^{(k)} (\delta^{(k)})^T\,.
\label{eq:combined_cov}
\end{equation}
The eigendecomposition of $\Sigma_{\rm comb}$ yields modes spanning both types of uncertainty.
Each systematic adds at most one significant eigenmode, so an analysis with $K_{\rm sys}$ systematic sources and $k_{\rm stat}$ statistical modes requires at most $k_{\rm stat} + K_{\rm sys}$ modes, compared to $N + K_{\rm sys}$ parameters in the histogram approach.

In the limit where statistical uncertainty is negligible ($\Sigma_{\rm stat} \to 0$), the combined covariance reduces to the outer product $\delta\delta^T$ for a single systematic, and Eq.~\eqref{eq:gp_expected} reproduces the InterpCode~4 piecewise-exponential morphing of Eq.~\eqref{eq:interpcode4}.
The eigenmode framework therefore contains both Barlow--Beeston and HistFactory interpolation as special cases.

\section{Benchmark design}
\label{sec:benchmarks}

The GP template construction of Section~\ref{sec:gp} is tested against the standard histogram-based approach in two experiments designed to probe complementary regimes.
Both use the same \histimator{} inference stack (Sec.~\ref{sec:software}) and differ only in the template construction and constraint parameterisation.

\subsection{Experiment~A: statistically limited rare-resonance search}
\label{sec:bench_a}

Experiment~A mimics a diphoton resonance search at the sensitivity frontier.
The background follows a power-law-exponential form typical of ATLAS and CMS diphoton analyses~\cite{ATLAS:2012higgs}, with $N_{\rm bkg} = 200$ expected data events in the invariant mass range $105$--$160$~GeV\@.
The signal is a narrow Gaussian at $130$~GeV with width $\sigma_s = 2$~GeV (FWHM $\approx 4.7$~GeV) and $N_{\rm sig} = 10$ expected events.
Templates are constructed from a single MC sample that is subsampled to varying sizes $N_{\rm MC} \in \{100, 200, 500, 1000, 2000\}$, corresponding to MC-to-data luminosity ratios $\tau \in
\{0.5, 1, 2.5, 5, 10\}$.
Figure~\ref{fig:variance_bound} shows the truth distribution and the GP and histogram templates at one representative $N_{\rm MC}=500$ draw.

The experiment addresses two questions.
The first is the binning dilemma: when MC statistics are limited, the analyst must choose between coarse bins, which give adequate MC counts per bin at the cost of smearing the signal, and fine bins, which resolve the signal at the cost of leaving many bins with few MC events.
Unequal bins may also be employed to remedy this however this choice will distort the shape of the distribution.
Furthermore is typical for distributions to be oversampled in MC, and analysts are encouraged to choose a binning that represents the resolution of the detector measurement rather than that of the template.
The second is template stability: the rate at which the background template degrades as $N_{\rm MC}$ decreases, and whether the GP maintains usable uncertainty quantification at MC budgets that break the histogram.

\subsection{Experiment~B: systematically limited two-channel
  measurement}
\label{sec:bench_b}

\begin{figure}[t]
\centering
\includegraphics[width=\columnwidth]{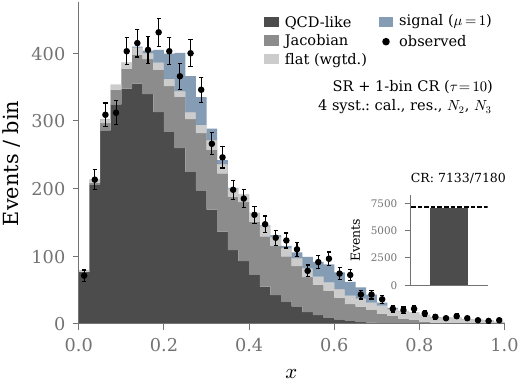}
\caption{Experiment~B truth distribution: three background components (QCD-like, Jacobian peak, flat/weighted) stacked with the signal mixture on the BDT discriminant $x\in[0,1]$.
The signal-to-background ratio in the SR is approximately $7\%$.
4 systematic variations are considered: the calibration shift, resolution smearing, normalisation on the Jacobian background ($N_2$), and the normalisation on the flat background $N_3$}
\label{fig:datasets}
\end{figure}

Experiment~B mimics a precision cross-section measurement with multiple background components, a control region, and several sources of systematic uncertainty.
Figure~\ref{fig:datasets} shows the Signal Region (SR) truth distribution.
Three backgrounds populate the SR: a steeply falling QCD-like component ($N_{\rm bkg1} = 4000$), a component with a Jacobian peak at $x \approx 0.2$ ($N_{\rm bkg2} = 1500$), and a minor flat component ($N_{\rm bkg3} = 400$) constructed from weighted MC events to test non-unit-weight handling.
The signal is a Gaussian mixture with $N_{\rm sig} = 400$ events, giving a signal-to-background ratio of approximately 7\% in the SR\@.
A single-bin control region (CR)
constrains the dominant background normalisation.  MC samples are
ten times the data ($\tau = 10$).

Four systematic uncertainties are included.  A calibration shift ($\pm 3$\% on backgrounds 1 and 2) moves events between bins.
A resolution smearing ($\pm 5$\% width change on background 1) broadens and narrows peaks.
Two normalisation uncertainties ($\pm 15$\% on backgrounds 2 and 3) adjust the relative composition.
The combined effect is that systematic uncertainty dominates MC statistical uncertainty in the bulk of the distribution, with the ratio exceeding 5:1 at intermediate $x$.

The GP eigenmode method fits a GP to the combined SR template histogram and extracts the leading eigenmodes of the combined (stat~+~systematic) covariance of Eq.~\eqref{eq:combined_cov}.
The histogram method uses Barlow--Beeston gammas for all bins plus explicit histosys and normsys modifiers.
Both methods share the same data, the same signal template, and the same control region constraint.

\section{Experiment~A: statistically limited search}
\label{sec:exp_a}

\subsection{The binning dilemma}

\begin{figure*}[t]
\centering
\includegraphics[width=\textwidth]{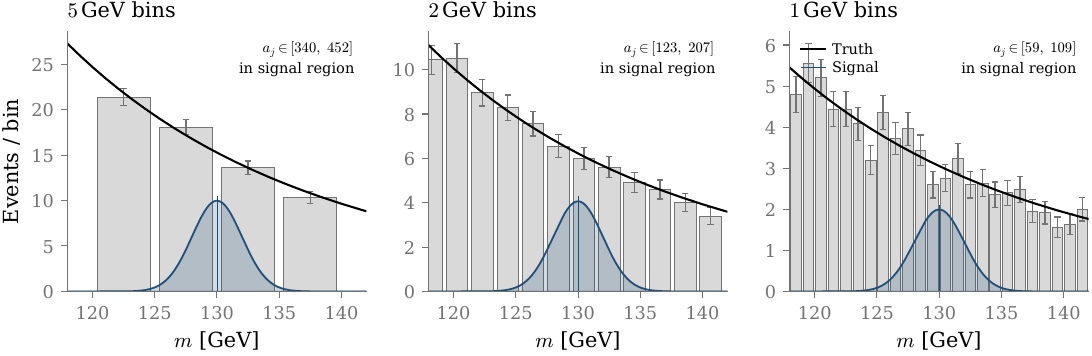}
\caption{The binning dilemma for the Experiment~A rare-resonance search.
The same $N_{\rm MC} = 500$ background events are binned at $5$~GeV (\textbf{left}), $2$~GeV (\textbf{centre}), and $1$~GeV (\textbf{right}).
Blue bars show the MC histogram with $1/\sqrt{a_j}$ Poisson error bars; the black curve is the truth; the shaded peak is the signal (scaled for visibility).
Annotations report the MC count range per bin in the signal region.
At coarse binning the signal is smeared; at fine binning the template is noisy.}
\label{fig:binning}
\end{figure*}

Figure~\ref{fig:binning} illustrates the tension between signal resolution and template precision that arises when the MC sample is small. 
The same $N_{\rm MC} = 500$ events ($\tau = 2.5$) are binned at three different widths.
At $5$~GeV per bin, each bin contains $18$--$46$ MC events, giving per-bin Poisson uncertainties of $15$--$23$\%.
The signal, however, is smeared across one to two bins and is essentially unresolvable from a background fluctuation.
At $2$~GeV per bin, the signal peak is resolved over two to three bins, but template precision degrades: bins in the signal region contain $5$--$18$ MC events, with uncertainties of $23$--$45$\%.
At $1$~GeV per bin, the signal shape is cleanly resolved, but several bins contain fewer than $5$ MC events.
Any histogram-based analysis in this last regime requires bin merging or vertical interpolation, defeating the purpose of the fine binning.

The GP template resolves this tension.
Fitted to the same MC sample, it reconstructs the smooth background shape with per-bin posterior uncertainty of $8$--$15$\% across the full range, roughly half the histogram uncertainty at comparable bin centres.
The signal remains resolvable against this smoother background regardless of how the underlying MC counts are binned for the GP fit, because the posterior is evaluated on an arbitrary grid after fitting.

\subsection{Template stability across MC budget}

The more quantitative question is how the template quality degrades as $N_{\rm MC}$ decreases.
The GP and histogram templates are constructed from a common underlying MC sample, progressively subsampled to $N_{\rm MC} = 2000, 500, 200, 100$.
At each subsample size, the GP is refitted with hyperparameters re-optimised via the marginal likelihood.

At $N_{\rm MC} = 2000$ ($\tau = 10$), both templates are close to truth and the uncertainties are small.
At $N_{\rm MC} = 500$ ($\tau = 2.5$), the histogram in the signal region ($125$--$135$~GeV) shows $1/\sqrt{a_j}$ uncertainties of $20$--$35$\%, while the GP posterior band is $8$--$12$\%.
At $N_{\rm MC} = 200$ ($\tau = 1$), several histogram bins in the signal region have fewer than $5$ MC events and the uncertainty exceeds $45$\%; the GP posterior remains at $12$--$18$\% because the smoothness prior pools information across the full domain.
At $N_{\rm MC} = 100$ ($\tau = 0.5$), the histogram is unusable in the tails, with empty bins appearing above $145$~GeV, while the GP provides a degraded but continuous estimate across the full range.

This behaviour follows directly from the variance bound of Eq.~\eqref{eq:variance_bound}: the GP posterior variance is bounded above by $1/a_j$ but the smoothness prior makes it strictly smaller wherever the kernel enforces correlation.
The practical consequence is that the GP template remains viable at MC budgets where the histogram template would require either aggressive bin merging or a substantial increase in MC production.

Template stability is a property of the posterior covariance and does not, on its own, fix the parameter of interest.
We test what the posterior implies for $\hat\mu$ through a closed-loop pseudo-experiment study at $\tau \in \{10, 5, 2.5, 1, 0.5\}$, in which $2{,}000$ MC subsamples per $\tau$ are drawn from a common pool, each subsample is used to construct both a GP and a Barlow--Beeston template, and each template is fitted to the same fixed observed dataset.
At $\tau = 10$ the two methods are quantitatively matched: BB has pull mean $-0.97$ and pull width $0.21$ with $68\%$ interval coverage of $56\%$, against GP pull mean $-0.96$, pull width $0.17$, and coverage $57\%$.
Both methods undercover at fixed $\tau=10$ because the rare-resonance signal is partially absorbed by the per-bin nuisance parameters, the same finite-MC effect quantified by Alexe \textit{et
al.}~\cite{Alexe:2024undercoverage} for high-statistics counting experiments.
As $\tau$ decreases, the BB pull width inflates and the BB bias attenuates, reaching width $1.40$ and bias $+0.10$ at $\tau=0.5$.
This reflects the increasingly loose BB auxiliary constraint that allows the per-bin gammas to track the data.
The GP pull width remains close to its high-MC value of $0.45$--$0.49$ while the GP bias drifts in the opposite direction to $-0.73$ at $\tau=0.5$.
This is the bias--variance signature of the two-step plug-in versus joint-profile distinction analysed in Appendix~\ref{app:semiparametric}: at small $\tau$ the joint
profile relaxes its template constraint and approaches unbiasedness with widened intervals, while the plug-in retains a tight posterior at the price of a residual smoothing bias that scales with the projection of the template error onto the signal direction.

\section{Experiment~B: systematically limited measurement}
\label{sec:exp_b}

\subsection{Eigenmode decomposition}

\begin{figure*}[t]
\centering
\includegraphics[width=\textwidth]{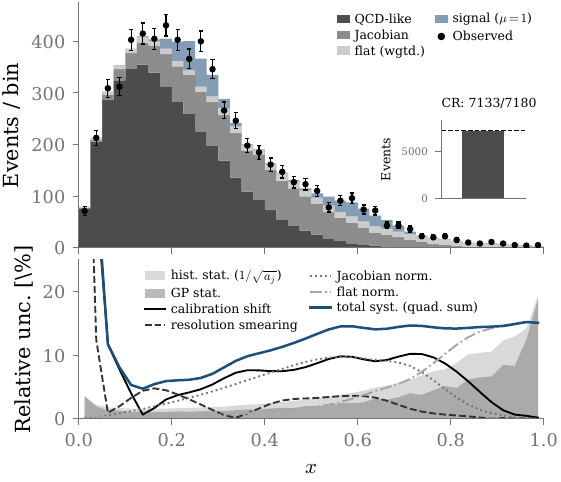}
\caption{\textbf{Top:} SR yields, stacked from the three background components and the $\mu=1$ signal, with observed counts overlaid as filled circles; the inset shows the single-bin control region
(observed versus expected total).
\textbf{Bottom:} per-bin contribution to the SR template uncertainty, plotted as a percentage of the nominal expected yield.
The two grey shaded bands are the statistical-only uncertainties of the two template representations: the histogram $1/\sqrt{a_j}$ (light grey) and the GP posterior standard deviation in log-rate space (mid grey).
The four line styles are the GP-smoothed shape systematics:
solid black for the calibration shift,
dashed for the resolution smearing,
dotted for the Jacobian-peak normalisation, 
dash-dotted for the flat-component normalisation.
The thick navy curve is the quadrature sum of the four systematics.
The eigenmode decomposition of Eq.~\eqref{eq:combined_cov} replaces these directions and the histogram per-bin gammas with a handful of amplitudes; the leading mode shapes are shown in Appendix~\ref{app:eigenstructure} (Fig.~\ref{fig:mode_shapes}).}
\label{fig:eigenmodes}
\end{figure*}

Figure~\ref{fig:eigenmodes} shows the eigenmode decomposition of the combined (stat~+~systematic) covariance for the Experiment~B SR template.
Panel~(a) displays the nominal GP template with the combined systematic envelope, computed as the quadrature sum of the four systematic directions of Eq.~\eqref{eq:sys_direction}.
The calibration and resolution systematics dominate at high $x$ where the steeply falling QCD background amplifies small shifts, while the normalisation uncertainties contribute uniformly.

Panel~(b) shows the four leading eigenmodes of $\Sigma_{\rm comb}$ as $\pm 1\sigma$ template deformations.
The leading mode ($\tilde\lambda_1 \approx 2.5$) captures the calibration-shift direction.
Modes~2--4 correspond to a mixture of the resolution smearing and the two normalisation uncertainties, with eigenvalues of order $0.2$--$0.4$.
The eigenvalue spectrum decays smoothly from mode~5 onward at a rate set by the GP statistical posterior on the $40$-bin grid.
The cumulative variance fraction reaches 95\% at $k = 6$, compared to $44$ parameters ($40$ gammas $+ 4$ histosys/normsys nuisance parameters) required by the histogram approach.
The compression ratio at this variance threshold is approximately $7$:$1$; tightening the threshold to 99\% raises the mode count to $k = 11$, still well below the histogram budget.

\subsection{Linearity and coverage}

Pseudo-experiments are generated at four injected signal strengths
$\mu_{\rm true} \in \{0, 0.5, 1.0, 2.0\}$, with observed data
Poisson-fluctuated around the corresponding expected counts.  For
each pseudo-experiment, both the GP eigenmode and Hist+BB models are
fitted, and $\hat\mu$ together with its parabolic (Hesse) error
$\sigma_{\hat\mu}$ is recorded.  The full toy ensemble of $2{,}000$
pseudo-experiments per injection point ($8{,}000$ total per method)
is generated in parallel through the
\texttt{run\_pseudo\_experiments} interface of \histimator{}
(Sec.~\ref{sec:software}).

\begin{figure}[t]
\centering
\includegraphics[width=\columnwidth]{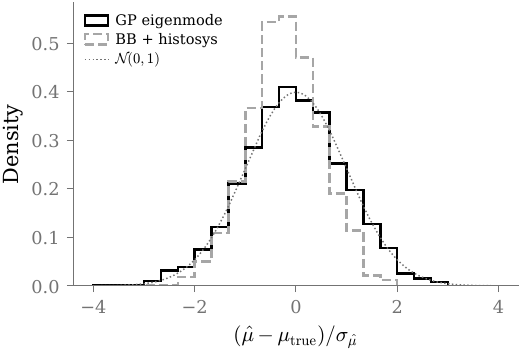}
\caption{Pull distribution at $\mu_{\rm true}=1$ for both methods across $2{,}000$ pseudo-experiments, against the unit Gaussian reference.
The GP eigenmode pull (solid) is centred at zero and unit-width whereas the Barlow--Beeston pull (dashed) does not match the Gaussian assumptions.}
\label{fig:validation}
\end{figure}

Bias, pull width, and interval coverage are summarised in Table~\ref{tab:exp_b_summary}.
The GP eigenmode method tracks the true signal strength with absolute bias below $0.006$ and pull width between $0.96$ and $0.99$ at every injection point, against the nominal unit width.
The Barlow--Beeston histogram method has bias of $\approx -0.04$ and pull width $0.65$--$0.73$.
The narrower BB pull reflects an over-estimation of $\sigma_{\hat\mu}$ in the rich nuisance space of $44$ correlated parameters, and is mirrored in the BB intervals: $68\%$ coverage of $81.7$--$85.5\%$ and $95\%$ coverage of $99.1$--$99.6\%$, both well above the nominal $68.3\%$ and $95.0\%$.
The eigenmode intervals sit at $67.7$--$70.5\%$ and $95.2$--$96.0\%$, consistent with nominal across all injection points, and are obtained directly from the parabolic Hesse error without a profile-likelihood scan.
Approximately $9\%$ of BB fits do not converge with the default strategy, against $0\%$ for the eigenmode parameterisation; the factor-of-six reduction in nuisance dimensionality translates directly into fit robustness.
Figure~\ref{fig:validation} shows the pull-shape comparison at $\mu_{\rm true}=1$.

\begin{table}[t]
\centering
\caption{Experiment~B validation summary across $2{,}000$
pseudo-experiments per $\mu_{\rm true}$.  Bias is
$\langle\hat\mu\rangle - \mu_{\rm true}$ with standard error on the
mean.  Pull width is the empirical standard deviation of
$(\hat\mu-\mu_{\rm true})/\sigma_{\hat\mu}$.  $C_{68}$ and $C_{95}$
are the fractions of pseudo-experiments with the parabolic
$1\sigma$ and $1.96\sigma$ interval covering $\mu_{\rm true}$;
nominal values are $68.3\%$ and $95.0\%$.  $n_{\rm ok}$ is the
number of converged fits.}
\label{tab:exp_b_summary}
\begin{tabular}{c c r r r r r}
\hline\hline
$\mu_{\rm true}$ & method & $n_{\rm ok}$ & bias & pull width & $C_{68}$ & $C_{95}$ \\
\hline
$0.0$ & GP & $2000$ & $-0.003(3)$ & $0.96$ & $70.5\%$ & $95.3\%$ \\
$0.0$ & BB & $1822$ & $-0.042(3)$ & $0.65$ & $85.5\%$ & $99.6\%$ \\
$0.5$ & GP & $2000$ & $+0.001(3)$ & $0.99$ & $67.7\%$ & $95.2\%$ \\
$0.5$ & BB & $1838$ & $-0.040(3)$ & $0.70$ & $83.0\%$ & $99.5\%$ \\
$1.0$ & GP & $2000$ & $-0.005(3)$ & $0.99$ & $68.5\%$ & $95.2\%$ \\
$1.0$ & BB & $1783$ & $-0.045(4)$ & $0.72$ & $81.8\%$ & $99.2\%$ \\
$2.0$ & GP & $2000$ & $-0.004(4)$ & $0.98$ & $68.7\%$ & $96.0\%$ \\
$2.0$ & BB & $1781$ & $-0.040(4)$ & $0.73$ & $81.7\%$ & $99.1\%$ \\
\hline\hline
\end{tabular}
\end{table}

\subsection{Diagnostics and pull interpretation}

An analyst adopting a new method needs to verify that standard
diagnostic tools still work.  The eigenmode basis trades per-bin
interpretability for compactness: a large pull on $z_3$ means ``the
third shape mode is shifted,'' which requires projection back onto
bins.  The resolution is straightforward.  The bin-level equivalent
pull at position $j$ is
\begin{equation}
p_j^{\rm proj} = \frac{\exp\!\left(\sum_{i=1}^{k}
\sqrt{\tilde\lambda_i}\, \hat{z}_i\, [\tilde{v}_i]_j\right) - 1}
{\sigma_j^{\rm BB}}\,,
\label{eq:projected_pull}
\end{equation}
where $\hat{z}_i$ are the fitted eigenmode amplitudes and
$\sigma_j^{\rm BB} = 1/\sqrt{a_j}$ is the BB reference scale.  This
projection recovers familiar per-bin pulls directly comparable to
BB gamma pulls.

\begin{figure*}[t]
\centering
\includegraphics[width=\textwidth]{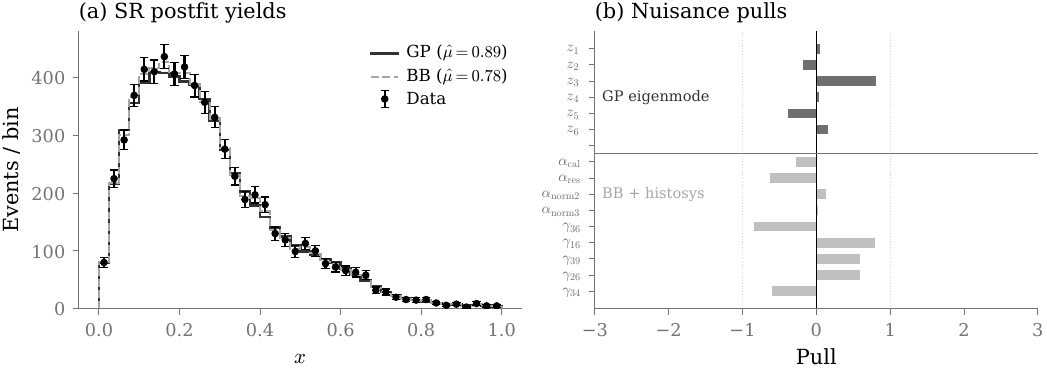}
\caption{Fit diagnostics for a single Experiment~B pseudo-experiment
at $\mu_{\rm true} = 1$.
\textbf{(a)}~Postfit yields: GP eigenmodes (solid) and Hist+BB
(dashed) both track the observed data (points with error bars).
Recovered $\hat\mu$ values are annotated.
\textbf{(b)}~Bin-level projected pulls from the eigenmode fit
(dark bars) compared with BB gamma pulls (light bars).}
\label{fig:diagnostics}
\end{figure*}

Figure~\ref{fig:diagnostics} demonstrates this for a representative
pseudo-experiment at $\mu_{\rm true} = 1$.  Panel~(a) shows postfit
yields in the SR, where both methods track the data.  Panel~(b)
compares the projected bin-level pulls from the eigenmode fit with the
BB gamma pulls.  The two patterns agree in the bulk of the
distribution, where the truncated eigenmode basis carries the
information that drives the fit.  At the bin edges of the spectrum
the BB gamma pulls remain near zero while the projected pulls retain
non-zero values of order unity: this is the expected behaviour of a
truncated reconstruction of an essentially per-bin random pattern,
where the eigenmodes that would absorb the edge fluctuations are below
the variance threshold and have been removed.  The eigenmode pulls
themselves provide the diagnostic information in a compressed,
importance-ordered form: a large $\hat z_1$ in the calibration
direction has a different physical meaning from a large $\hat z_4$
along a statistical mode, and the ordering is itself a part of the
diagnostic.

\section{Implementation}
\label{sec:software}

The method is implemented in \histimator{}, an open-source Python
package providing an imperative API for HistFactory-style template
likelihoods.  Channels, samples, and modifiers are constructed
programmatically; the GP template is selected by passing
\texttt{template\_type="gp"} when constructing a \texttt{Sample}, and
the eigenmode constraint by attaching an \texttt{EigenmodeConstraint}
or via the \texttt{add\_eigenmode\_staterror} convenience method.
The inference stack supports profile likelihood fitting via
\texttt{iminuit}~\cite{iminuit:2020}, asymptotic
CL$_s$~\cite{Read:2002,Cowan:2010js}, discovery significance, and
diagnostics.  The pseudo-experiment ensembles used for the closure
and coverage studies of Sections~\ref{sec:exp_a}
and~\ref{sec:exp_b} are run in parallel via the
\texttt{run\_pseudo\_experiments} interface, which dispatches across worker processes with stable
seed-ordered output and deterministic seeding.
The toy ensembles reported in this paper run in linear-time speed-up to roughly the
core count (e.g., $7$~min on $16$ cores for the $2{,}000$-PE,
five-$\tau$ Experiment~A closure that takes $\sim 1$~h serial).
Appendix~\ref{app:histfactory} validates the full modifier chain
and CL$_s$ against the standard histogram approach.  The package is
available at \url{https://github.com/histimator/histimator}; the
release accompanying this paper is
\histimator{}~v$0.4.0$~\cite{histimator:2025}.

\section{Conclusions and Discussions}
\label{sec:conclusions}

We have presented a method for representing statistical and systematic uncertainties in the construction of smooth templates for the HistFactory likelihood from Gaussian process posteriors fitted to binned Monte Carlo data.
The log-Gaussian Cox process provides a Poisson-native model for the MC template counts, and the Laplace approximation to the GP posterior yields both a smooth template (the posterior mean) and a structured uncertainty (the posterior covariance).
Systematic shape variations are incorporated through rank-1 covariance updates, and the eigendecomposition of the combined covariance produces a compact basis that encodes both statistical and systematic template uncertainty.

We have proved that this construction contains Barlow--Beeston as a
limiting case and that the GP posterior variance is bounded above by
the BB variance at every bin.  The eigenmode framework also recovers
HistFactory InterpCode~4 interpolation in the limit where statistical
uncertainty is negligible, so the standard modifier chain is subsumed
rather than replaced.

Two benchmark experiments demonstrate the method.  In the
statistically limited regime (Experiment~A), the GP template maintains
stable uncertainty quantification as the MC sample size decreases,
resolving the binning dilemma that forces histogram analyses into
coarse binning when MC statistics are poor.  In the systematically
limited regime (Experiment~B), the eigenmode method achieves
equivalent linearity, coverage, and pull calibration to the histogram
approach while compressing the template parameter space by an order of
magnitude and preserving the diagnostic power of per-bin pulls through
eigenmode projection.

The boundary of applicability of this construction is set by
semiparametric efficiency.  The Barlow--Beeston joint profile achieves
the semiparametric information bound for $\mu$ in the Poisson model;
the GP plug-in does not, because the smooth template is fitted in a
first step from the MC counts and not adapted to the data during
signal extraction.  Appendix~\ref{app:semiparametric} derives the
closed-form efficiency factor $\tau/(1{+}\tau)$ that quantifies this
gap and identifies the regime in which it matters.  At
$\tau=2.5$ (Experiment~A) approximately 29\% of the signal-strength
information is consumed by background uncertainty, and the joint
profile is structurally the right tool; at $\tau=10$ (Experiment~B)
the efficiency loss is below 9\% and the eigenmode compression is the
operationally important property.  The eigenmode count moreover
scales with the effective dimensionality of the GP kernel rather than
with the number of bins, so a three-dimensional template with $20^3$
bins requires of order $30$ eigenmode amplitudes against the
$8000$ Barlow--Beeston gammas, and the GP construction in addition
delivers the kernel-based covariance structure of the test-statistic
field needed for analytic look-elsewhere
estimates~\cite{Gross:2010qma,Gandrakota:2022gplee}, a capability
the histogram has not.  Exact GP inference scales as $O(N^3)$ in
the number of bins; approximate methods based on sparse
GPs~\cite{Titsias:2009}, the SPDE
approach~\cite{Lindgren:2011spde}, or structured kernel
interpolation~\cite{Wilson:2015ski} reduce this to near-linear cost
and are a natural next step.

The method is implemented in the open-source \histimator{} package
(Sec.~\ref{sec:software}).

\section*{Acknowledgments}
We thank the developers of
\texttt{iminuit}~\cite{iminuit:2020}
and the ROOT team~\cite{ROOT:2024} for software infrastructure on
which this work builds.


\bibliography{references}

\begin{thebibliography}{36}%
\makeatletter
\providecommand \@ifxundefined [1]{%
 \@ifx{#1\undefined}
}%
\providecommand \@ifnum [1]{%
 \ifnum #1\expandafter \@firstoftwo
 \else \expandafter \@secondoftwo
 \fi
}%
\providecommand \@ifx [1]{%
 \ifx #1\expandafter \@firstoftwo
 \else \expandafter \@secondoftwo
 \fi
}%
\providecommand \natexlab [1]{#1}%
\providecommand \enquote  [1]{``#1''}%
\providecommand \bibnamefont  [1]{#1}%
\providecommand \bibfnamefont [1]{#1}%
\providecommand \citenamefont [1]{#1}%
\providecommand \href@noop [0]{\@secondoftwo}%
\providecommand \href [0]{\begingroup \@sanitize@url \@href}%
\providecommand \@href[1]{\@@startlink{#1}\@@href}%
\providecommand \@@href[1]{\endgroup#1\@@endlink}%
\providecommand \@sanitize@url [0]{\catcode `\\12\catcode `\$12\catcode `\&12\catcode `\#12\catcode `\^12\catcode `\_12\catcode `\%12\relax}%
\providecommand \@@startlink[1]{}%
\providecommand \@@endlink[0]{}%
\providecommand \url  [0]{\begingroup\@sanitize@url \@url }%
\providecommand \@url [1]{\endgroup\@href {#1}{\urlprefix }}%
\providecommand \urlprefix  [0]{URL }%
\providecommand \Eprint [0]{\href }%
\providecommand \doibase [0]{https://doi.org/}%
\providecommand \selectlanguage [0]{\@gobble}%
\providecommand \bibinfo  [0]{\@secondoftwo}%
\providecommand \bibfield  [0]{\@secondoftwo}%
\providecommand \translation [1]{[#1]}%
\providecommand \BibitemOpen [0]{}%
\providecommand \bibitemStop [0]{}%
\providecommand \bibitemNoStop [0]{.\EOS\space}%
\providecommand \EOS [0]{\spacefactor3000\relax}%
\providecommand \BibitemShut  [1]{\csname bibitem#1\endcsname}%
\let\auto@bib@innerbib\@empty
\bibitem [{\citenamefont {Cranmer}\ \emph {et~al.}(2012)\citenamefont {Cranmer}, \citenamefont {Lewis}, \citenamefont {Moneta}, \citenamefont {Shibata},\ and\ \citenamefont {Verkerke}}]{Cranmer:2012sya}%
  \BibitemOpen
  \bibfield  {author} {\bibinfo {author} {\bibfnamefont {K.}~\bibnamefont {Cranmer}}, \bibinfo {author} {\bibfnamefont {G.}~\bibnamefont {Lewis}}, \bibinfo {author} {\bibfnamefont {L.}~\bibnamefont {Moneta}}, \bibinfo {author} {\bibfnamefont {A.}~\bibnamefont {Shibata}},\ and\ \bibinfo {author} {\bibfnamefont {W.}~\bibnamefont {Verkerke}},\ }\href {https://cds.cern.ch/record/1456844} {\emph {\bibinfo {title} {{HistFactory: A tool for creating statistical models for use with RooFit and RooStats}}}},\ \bibinfo {type} {Tech. Rep.}\ \bibinfo {number} {CERN-OPEN-2012-016}\ (\bibinfo {year} {2012})\BibitemShut {NoStop}%
\bibitem [{\citenamefont {Cowan}\ \emph {et~al.}(2011)\citenamefont {Cowan}, \citenamefont {Cranmer}, \citenamefont {Gross},\ and\ \citenamefont {Vitells}}]{Cowan:2010js}%
  \BibitemOpen
  \bibfield  {author} {\bibinfo {author} {\bibfnamefont {G.}~\bibnamefont {Cowan}}, \bibinfo {author} {\bibfnamefont {K.}~\bibnamefont {Cranmer}}, \bibinfo {author} {\bibfnamefont {E.}~\bibnamefont {Gross}},\ and\ \bibinfo {author} {\bibfnamefont {O.}~\bibnamefont {Vitells}},\ }\bibfield  {title} {\bibinfo {title} {{Asymptotic formulae for likelihood-based tests of new physics}},\ }\href {https://doi.org/10.1140/epjc/s10052-011-1554-0} {\bibfield  {journal} {\bibinfo  {journal} {Eur. Phys. J. C}\ }\textbf {\bibinfo {volume} {71}},\ \bibinfo {pages} {1554} (\bibinfo {year} {2011})},\ \Eprint {https://arxiv.org/abs/1007.1727} {arXiv:1007.1727} \BibitemShut {NoStop}%
\bibitem [{\citenamefont {{ATLAS Collaboration}}(2012)}]{ATLAS:2012higgs}%
  \BibitemOpen
  \bibfield  {author} {\bibinfo {author} {\bibnamefont {{ATLAS Collaboration}}},\ }\bibfield  {title} {\bibinfo {title} {{Observation of a new particle in the search for the Standard Model Higgs boson with the ATLAS detector at the LHC}},\ }\href {https://doi.org/10.1016/j.physletb.2012.08.020} {\bibfield  {journal} {\bibinfo  {journal} {Phys. Lett. B}\ }\textbf {\bibinfo {volume} {716}},\ \bibinfo {pages} {1} (\bibinfo {year} {2012})},\ \Eprint {https://arxiv.org/abs/1207.7214} {arXiv:1207.7214} \BibitemShut {NoStop}%
\bibitem [{\citenamefont {{CMS Collaboration}}(2012)}]{CMS:2012higgs}%
  \BibitemOpen
  \bibfield  {author} {\bibinfo {author} {\bibnamefont {{CMS Collaboration}}},\ }\bibfield  {title} {\bibinfo {title} {{Observation of a new boson at a mass of 125 GeV with the CMS experiment at the LHC}},\ }\href {https://doi.org/10.1016/j.physletb.2012.08.021} {\bibfield  {journal} {\bibinfo  {journal} {Phys. Lett. B}\ }\textbf {\bibinfo {volume} {716}},\ \bibinfo {pages} {30} (\bibinfo {year} {2012})},\ \Eprint {https://arxiv.org/abs/1207.7235} {arXiv:1207.7235} \BibitemShut {NoStop}%
\bibitem [{\citenamefont {{ATLAS Collaboration}}(2022)}]{ATLAS:2022higgs}%
  \BibitemOpen
  \bibfield  {author} {\bibinfo {author} {\bibnamefont {{ATLAS Collaboration}}},\ }\bibfield  {title} {\bibinfo {title} {{A detailed map of Higgs boson interactions by the ATLAS experiment ten years after the discovery}},\ }\href {https://doi.org/10.1038/s41586-022-04893-w} {\bibfield  {journal} {\bibinfo  {journal} {Nature}\ }\textbf {\bibinfo {volume} {607}},\ \bibinfo {pages} {52} (\bibinfo {year} {2022})}\BibitemShut {NoStop}%
\bibitem [{\citenamefont {Croft}\ and\ \citenamefont {{Histimator Collaboration}}(2026)}]{histimator:2025}%
  \BibitemOpen
  \bibfield  {author} {\bibinfo {author} {\bibfnamefont {V.}~\bibnamefont {Croft}}\ and\ \bibinfo {author} {\bibnamefont {{Histimator Collaboration}}},\ }\href {https://github.com/histimator/histimator} {\bibinfo {title} {{\texttt{Histimator}: Gaussian process templates for HistFactory likelihoods}}} (\bibinfo {year} {2026}),\ \bibinfo {note} {adds parallel pseudo-experiment driver \texttt{run\_pseudo\_experiments}, projected-pull diagnostic, ShapeFactor and InterpCode-4 limit tests.}\BibitemShut {Stop}%
\bibitem [{\citenamefont {Barlow}\ and\ \citenamefont {Beeston}(1993)}]{Barlow:1993dm}%
  \BibitemOpen
  \bibfield  {author} {\bibinfo {author} {\bibfnamefont {R.}~\bibnamefont {Barlow}}\ and\ \bibinfo {author} {\bibfnamefont {C.}~\bibnamefont {Beeston}},\ }\bibfield  {title} {\bibinfo {title} {{Fitting using finite Monte Carlo samples}},\ }\href {https://doi.org/10.1016/0010-4655(93)90005-W} {\bibfield  {journal} {\bibinfo  {journal} {Comput. Phys. Commun.}\ }\textbf {\bibinfo {volume} {77}},\ \bibinfo {pages} {219} (\bibinfo {year} {1993})}\BibitemShut {NoStop}%
\bibitem [{\citenamefont {Conway}(2011)}]{Conway:2011in}%
  \BibitemOpen
  \bibfield  {author} {\bibinfo {author} {\bibfnamefont {J.~S.}\ \bibnamefont {Conway}},\ }\bibfield  {title} {\bibinfo {title} {{Incorporating Nuisance Parameters in Likelihoods for Multisource Spectra}},\ }in\ \href@noop {} {\emph {\bibinfo {booktitle} {PHYSTAT 2011}}}\ (\bibinfo {year} {2011})\ \Eprint {https://arxiv.org/abs/1103.0354} {arXiv:1103.0354 [physics.data-an]} \BibitemShut {NoStop}%
\bibitem [{\citenamefont {Alexe}\ \emph {et~al.}(2024)\citenamefont {Alexe}, \citenamefont {Bendavid}, \citenamefont {Bianchini},\ and\ \citenamefont {Bruschini}}]{Alexe:2024undercoverage}%
  \BibitemOpen
  \bibfield  {author} {\bibinfo {author} {\bibfnamefont {B.}~\bibnamefont {Alexe}}, \bibinfo {author} {\bibfnamefont {J.}~\bibnamefont {Bendavid}}, \bibinfo {author} {\bibfnamefont {L.}~\bibnamefont {Bianchini}},\ and\ \bibinfo {author} {\bibfnamefont {A.}~\bibnamefont {Bruschini}},\ }\bibfield  {title} {\bibinfo {title} {{Under-coverage in high-statistics counting experiments with finite MC samples}},\ }\href@noop {} {\  (\bibinfo {year} {2024})},\ \Eprint {https://arxiv.org/abs/2401.10542} {arXiv:2401.10542 [physics.data-an]} \BibitemShut {NoStop}%
\bibitem [{\citenamefont {{M\o{}ller}}\ \emph {et~al.}(1998)\citenamefont {{M\o{}ller}}, \citenamefont {Syversveen},\ and\ \citenamefont {Waagepetersen}}]{Moller:1998lgcp}%
  \BibitemOpen
  \bibfield  {author} {\bibinfo {author} {\bibfnamefont {J.}~\bibnamefont {{M\o{}ller}}}, \bibinfo {author} {\bibfnamefont {A.~R.}\ \bibnamefont {Syversveen}},\ and\ \bibinfo {author} {\bibfnamefont {R.~P.}\ \bibnamefont {Waagepetersen}},\ }\bibfield  {title} {\bibinfo {title} {{Log Gaussian Cox Processes}},\ }\href {https://doi.org/10.1111/1467-9469.00115} {\bibfield  {journal} {\bibinfo  {journal} {Scand. J. Statist.}\ }\textbf {\bibinfo {volume} {25}},\ \bibinfo {pages} {451} (\bibinfo {year} {1998})}\BibitemShut {NoStop}%
\bibitem [{\citenamefont {Diggle}\ \emph {et~al.}(2013)\citenamefont {Diggle}, \citenamefont {Moraga}, \citenamefont {Rowlingson},\ and\ \citenamefont {Taylor}}]{Diggle:2013}%
  \BibitemOpen
  \bibfield  {author} {\bibinfo {author} {\bibfnamefont {P.~J.}\ \bibnamefont {Diggle}}, \bibinfo {author} {\bibfnamefont {P.}~\bibnamefont {Moraga}}, \bibinfo {author} {\bibfnamefont {B.}~\bibnamefont {Rowlingson}},\ and\ \bibinfo {author} {\bibfnamefont {B.~M.}\ \bibnamefont {Taylor}},\ }\href {https://doi.org/10.1214/13-STS441} {\emph {\bibinfo {title} {{Spatial and Spatio-Temporal Log-Gaussian Cox Processes: Extending the Geostatistical Paradigm}}}},\ Vol.~\bibinfo {volume} {28}\ (\bibinfo {year} {2013})\ pp.\ \bibinfo {pages} {542--563}\BibitemShut {NoStop}%
\bibitem [{\citenamefont {Frate}\ \emph {et~al.}(2017)\citenamefont {Frate}, \citenamefont {Cranmer}, \citenamefont {Kalia}, \citenamefont {Vandenberg-Rodes},\ and\ \citenamefont {Whiteson}}]{Frate:2017mai}%
  \BibitemOpen
  \bibfield  {author} {\bibinfo {author} {\bibfnamefont {M.}~\bibnamefont {Frate}}, \bibinfo {author} {\bibfnamefont {K.}~\bibnamefont {Cranmer}}, \bibinfo {author} {\bibfnamefont {S.}~\bibnamefont {Kalia}}, \bibinfo {author} {\bibfnamefont {A.}~\bibnamefont {Vandenberg-Rodes}},\ and\ \bibinfo {author} {\bibfnamefont {D.}~\bibnamefont {Whiteson}},\ }\bibfield  {title} {\bibinfo {title} {{Modeling Smooth Backgrounds and Generic Localized Signals with Gaussian Processes}},\ }\href@noop {} {\  (\bibinfo {year} {2017})},\ \Eprint {https://arxiv.org/abs/1709.05681} {arXiv:1709.05681 [physics.data-an]} \BibitemShut {NoStop}%
\bibitem [{\citenamefont {Gandrakota}\ \emph {et~al.}(2023)\citenamefont {Gandrakota}, \citenamefont {Lath}, \citenamefont {Morozov},\ and\ \citenamefont {Murthy}}]{Gandrakota:2023gph}%
  \BibitemOpen
  \bibfield  {author} {\bibinfo {author} {\bibfnamefont {A.}~\bibnamefont {Gandrakota}}, \bibinfo {author} {\bibfnamefont {A.}~\bibnamefont {Lath}}, \bibinfo {author} {\bibfnamefont {A.~V.}\ \bibnamefont {Morozov}},\ and\ \bibinfo {author} {\bibfnamefont {S.}~\bibnamefont {Murthy}},\ }\bibfield  {title} {\bibinfo {title} {{Model selection and signal extraction using Gaussian Process regression}},\ }\href {https://doi.org/10.1007/JHEP02(2023)230} {\bibfield  {journal} {\bibinfo  {journal} {JHEP}\ }\textbf {\bibinfo {volume} {02}},\ \bibinfo {pages} {230}}\BibitemShut {NoStop}%
\bibitem [{\citenamefont {Frid}\ \emph {et~al.}(2025)\citenamefont {Frid}, \citenamefont {Barak}, \citenamefont {Jairam}, \citenamefont {Kagan},\ and\ \citenamefont {Hyneman}}]{Frid:2025lgcp}%
  \BibitemOpen
  \bibfield  {author} {\bibinfo {author} {\bibfnamefont {I.}~\bibnamefont {Frid}}, \bibinfo {author} {\bibfnamefont {L.}~\bibnamefont {Barak}}, \bibinfo {author} {\bibfnamefont {S.}~\bibnamefont {Jairam}}, \bibinfo {author} {\bibfnamefont {M.}~\bibnamefont {Kagan}},\ and\ \bibinfo {author} {\bibfnamefont {R.}~\bibnamefont {Hyneman}},\ }\bibfield  {title} {\bibinfo {title} {{Log Gaussian Cox Process Background Modeling in High Energy Physics}},\ }\href@noop {} {\  (\bibinfo {year} {2025})},\ \Eprint {https://arxiv.org/abs/2508.11740} {arXiv:2508.11740 [hep-ex]} \BibitemShut {NoStop}%
\bibitem [{\citenamefont {Whitehorn}\ \emph {et~al.}(2013)\citenamefont {Whitehorn}, \citenamefont {van Santen},\ and\ \citenamefont {Boeser}}]{Whitehorn:2013pspline}%
  \BibitemOpen
  \bibfield  {author} {\bibinfo {author} {\bibfnamefont {N.}~\bibnamefont {Whitehorn}}, \bibinfo {author} {\bibfnamefont {J.}~\bibnamefont {van Santen}},\ and\ \bibinfo {author} {\bibfnamefont {S.}~\bibnamefont {Boeser}},\ }\bibfield  {title} {\bibinfo {title} {{Penalized splines for smooth representation of high-dimensional Monte Carlo datasets}},\ }\href {https://doi.org/10.1016/j.cpc.2013.04.008} {\bibfield  {journal} {\bibinfo  {journal} {Comput. Phys. Commun.}\ }\textbf {\bibinfo {volume} {184}},\ \bibinfo {pages} {2214} (\bibinfo {year} {2013})}\BibitemShut {NoStop}%
\bibitem [{\citenamefont {Lindgren}\ \emph {et~al.}(2011)\citenamefont {Lindgren}, \citenamefont {Rue},\ and\ \citenamefont {{Lindstr\"om}}}]{Lindgren:2011spde}%
  \BibitemOpen
  \bibfield  {author} {\bibinfo {author} {\bibfnamefont {F.}~\bibnamefont {Lindgren}}, \bibinfo {author} {\bibfnamefont {H.}~\bibnamefont {Rue}},\ and\ \bibinfo {author} {\bibfnamefont {J.}~\bibnamefont {{Lindstr\"om}}},\ }\bibfield  {title} {\bibinfo {title} {{An explicit link between Gaussian fields and Gaussian Markov random fields: the stochastic partial differential equation approach}},\ }\href {https://doi.org/10.1111/j.1467-9868.2011.00777.x} {\bibfield  {journal} {\bibinfo  {journal} {J. R. Stat. Soc. B}\ }\textbf {\bibinfo {volume} {73}},\ \bibinfo {pages} {423} (\bibinfo {year} {2011})}\BibitemShut {NoStop}%
\bibitem [{\citenamefont {Cranmer}(2001)}]{Cranmer:2000du}%
  \BibitemOpen
  \bibfield  {author} {\bibinfo {author} {\bibfnamefont {K.~S.}\ \bibnamefont {Cranmer}},\ }\bibfield  {title} {\bibinfo {title} {{Kernel estimation in high-energy physics}},\ }\href {https://doi.org/10.1016/S0010-4655(00)00243-5} {\bibfield  {journal} {\bibinfo  {journal} {Comput. Phys. Commun.}\ }\textbf {\bibinfo {volume} {136}},\ \bibinfo {pages} {198} (\bibinfo {year} {2001})}\BibitemShut {NoStop}%
\bibitem [{\citenamefont {Dembinski}\ and\ \citenamefont {{Abdelmotteleb}}(2022)}]{Dembinski:2022template}%
  \BibitemOpen
  \bibfield  {author} {\bibinfo {author} {\bibfnamefont {H.~P.}\ \bibnamefont {Dembinski}}\ and\ \bibinfo {author} {\bibfnamefont {A.~S.~W.}\ \bibnamefont {{Abdelmotteleb}}},\ }\bibfield  {title} {\bibinfo {title} {{A new maximum-likelihood method for template fits}},\ }\href {https://doi.org/10.1140/epjc/s10052-022-11019-z} {\bibfield  {journal} {\bibinfo  {journal} {Eur. Phys. J. C}\ }\textbf {\bibinfo {volume} {82}},\ \bibinfo {pages} {1043} (\bibinfo {year} {2022})}\BibitemShut {NoStop}%
\bibitem [{\citenamefont {Rasmussen}\ and\ \citenamefont {Williams}(2006)}]{RW:2006}%
  \BibitemOpen
  \bibfield  {author} {\bibinfo {author} {\bibfnamefont {C.~E.}\ \bibnamefont {Rasmussen}}\ and\ \bibinfo {author} {\bibfnamefont {C.~K.~I.}\ \bibnamefont {Williams}},\ }\href@noop {} {\emph {\bibinfo {title} {{Gaussian Processes for Machine Learning}}}}\ (\bibinfo  {publisher} {MIT Press},\ \bibinfo {year} {2006})\BibitemShut {NoStop}%
\bibitem [{\citenamefont {de~Boor}(2001)}]{deBoor:2001}%
  \BibitemOpen
  \bibfield  {author} {\bibinfo {author} {\bibfnamefont {C.}~\bibnamefont {de~Boor}},\ }\href@noop {} {\emph {\bibinfo {title} {{A Practical Guide to Splines}}}},\ \bibinfo {edition} {revised}\ ed.\ (\bibinfo  {publisher} {Springer},\ \bibinfo {year} {2001})\BibitemShut {NoStop}%
\bibitem [{\citenamefont {Eilers}\ and\ \citenamefont {Marx}(1996)}]{Eilers:1996pspline}%
  \BibitemOpen
  \bibfield  {author} {\bibinfo {author} {\bibfnamefont {P.~H.~C.}\ \bibnamefont {Eilers}}\ and\ \bibinfo {author} {\bibfnamefont {B.~D.}\ \bibnamefont {Marx}},\ }\bibfield  {title} {\bibinfo {title} {{Flexible Smoothing with B-splines and Penalties}},\ }\href@noop {} {\bibfield  {journal} {\bibinfo  {journal} {Stat. Sci.}\ }\textbf {\bibinfo {volume} {11}},\ \bibinfo {pages} {89} (\bibinfo {year} {1996})}\BibitemShut {NoStop}%
\bibitem [{\citenamefont {Kimeldorf}\ and\ \citenamefont {Wahba}(1970)}]{Kimeldorf:1970}%
  \BibitemOpen
  \bibfield  {author} {\bibinfo {author} {\bibfnamefont {G.}~\bibnamefont {Kimeldorf}}\ and\ \bibinfo {author} {\bibfnamefont {G.}~\bibnamefont {Wahba}},\ }\bibfield  {title} {\bibinfo {title} {{A Correspondence Between Bayesian Estimation on Stochastic Processes and Smoothing by Splines}},\ }\href {https://doi.org/10.1214/aoms/1177697089} {\bibfield  {journal} {\bibinfo  {journal} {Ann. Math. Stat.}\ }\textbf {\bibinfo {volume} {41}},\ \bibinfo {pages} {495} (\bibinfo {year} {1970})}\BibitemShut {NoStop}%
\bibitem [{\citenamefont {Dembinski}\ \emph {et~al.}(2020)\citenamefont {Dembinski} \emph {et~al.}}]{iminuit:2020}%
  \BibitemOpen
  \bibfield  {author} {\bibinfo {author} {\bibfnamefont {H.~P.}\ \bibnamefont {Dembinski}} \emph {et~al.},\ }\href {https://doi.org/10.5281/zenodo.3949207} {\bibinfo {title} {{\texttt{iminuit}: A Python interface to MINUIT}}} (\bibinfo {year} {2020})\BibitemShut {NoStop}%
\bibitem [{\citenamefont {Read}(2002)}]{Read:2002}%
  \BibitemOpen
  \bibfield  {author} {\bibinfo {author} {\bibfnamefont {A.~L.}\ \bibnamefont {Read}},\ }\bibfield  {title} {\bibinfo {title} {{Presentation of search results: The CL$_s$ technique}},\ }\href {https://doi.org/10.1088/0954-3899/28/10/313} {\bibfield  {journal} {\bibinfo  {journal} {J. Phys. G}\ }\textbf {\bibinfo {volume} {28}},\ \bibinfo {pages} {2693} (\bibinfo {year} {2002})}\BibitemShut {NoStop}%
\bibitem [{\citenamefont {Gross}\ and\ \citenamefont {Vitells}(2010)}]{Gross:2010qma}%
  \BibitemOpen
  \bibfield  {author} {\bibinfo {author} {\bibfnamefont {E.}~\bibnamefont {Gross}}\ and\ \bibinfo {author} {\bibfnamefont {O.}~\bibnamefont {Vitells}},\ }\bibfield  {title} {\bibinfo {title} {{Trial factors for the look elsewhere effect in high energy physics}},\ }\href@noop {} {\bibfield  {journal} {\bibinfo  {journal} {Eur. Phys. J. C}\ }\textbf {\bibinfo {volume} {70}},\ \bibinfo {pages} {525} (\bibinfo {year} {2010})},\ \Eprint {https://arxiv.org/abs/1005.1891} {arXiv:1005.1891 [hep-ex]} \BibitemShut {NoStop}%
\bibitem [{\citenamefont {Gandrakota}\ \emph {et~al.}(2022)\citenamefont {Gandrakota}, \citenamefont {Cueto},\ and\ \citenamefont {Gross}}]{Gandrakota:2022gplee}%
  \BibitemOpen
  \bibfield  {author} {\bibinfo {author} {\bibfnamefont {A.}~\bibnamefont {Gandrakota}}, \bibinfo {author} {\bibfnamefont {A.}~\bibnamefont {Cueto}},\ and\ \bibinfo {author} {\bibfnamefont {E.}~\bibnamefont {Gross}},\ }\bibfield  {title} {\bibinfo {title} {{GP-based estimation of the look-elsewhere effect}},\ }\href@noop {} {\  (\bibinfo {year} {2022})},\ \Eprint {https://arxiv.org/abs/2206.12328} {arXiv:2206.12328 [physics.data-an]} \BibitemShut {NoStop}%
\bibitem [{\citenamefont {Titsias}(2009)}]{Titsias:2009}%
  \BibitemOpen
  \bibfield  {author} {\bibinfo {author} {\bibfnamefont {M.~K.}\ \bibnamefont {Titsias}},\ }\bibfield  {title} {\bibinfo {title} {{Variational Learning of Inducing Variables in Sparse Gaussian Processes}},\ }\bibfield  {booktitle} {\emph {\bibinfo {booktitle} {AISTATS 2009}},\ }\href@noop {} {\bibfield  {journal} {\bibinfo  {journal} {Proc. Mach. Learn. Res.}\ }\textbf {\bibinfo {volume} {5}},\ \bibinfo {pages} {567} (\bibinfo {year} {2009})}\BibitemShut {NoStop}%
\bibitem [{\citenamefont {Wilson}\ and\ \citenamefont {Nickisch}(2015)}]{Wilson:2015ski}%
  \BibitemOpen
  \bibfield  {author} {\bibinfo {author} {\bibfnamefont {A.~G.}\ \bibnamefont {Wilson}}\ and\ \bibinfo {author} {\bibfnamefont {H.}~\bibnamefont {Nickisch}},\ }\bibfield  {title} {\bibinfo {title} {{Kernel Interpolation for Scalable Structured Gaussian Processes (KISS-GP)}},\ }\bibfield  {booktitle} {\emph {\bibinfo {booktitle} {ICML 2015}},\ }\href@noop {} {\bibfield  {journal} {\bibinfo  {journal} {Proc. Mach. Learn. Res.}\ }\textbf {\bibinfo {volume} {37}},\ \bibinfo {pages} {1775} (\bibinfo {year} {2015})}\BibitemShut {NoStop}%
\bibitem [{\citenamefont {Brun}\ \emph {et~al.}(2024)\citenamefont {Brun}, \citenamefont {Rademakers} \emph {et~al.}}]{ROOT:2024}%
  \BibitemOpen
  \bibfield  {author} {\bibinfo {author} {\bibfnamefont {R.}~\bibnamefont {Brun}}, \bibinfo {author} {\bibfnamefont {F.}~\bibnamefont {Rademakers}}, \emph {et~al.},\ }\href {https://root.cern} {\bibinfo {title} {{ROOT: An Object-Oriented Data Analysis Framework}}} (\bibinfo {year} {2024})\BibitemShut {NoStop}%
\bibitem [{\citenamefont {Bickel}\ \emph {et~al.}(1993)\citenamefont {Bickel}, \citenamefont {Klaassen}, \citenamefont {Ritov},\ and\ \citenamefont {Wellner}}]{Bickel:1993}%
  \BibitemOpen
  \bibfield  {author} {\bibinfo {author} {\bibfnamefont {P.~J.}\ \bibnamefont {Bickel}}, \bibinfo {author} {\bibfnamefont {C.~A.~J.}\ \bibnamefont {Klaassen}}, \bibinfo {author} {\bibfnamefont {Y.}~\bibnamefont {Ritov}},\ and\ \bibinfo {author} {\bibfnamefont {J.~A.}\ \bibnamefont {Wellner}},\ }\href@noop {} {\emph {\bibinfo {title} {{Efficient and Adaptive Estimation for Semiparametric Models}}}}\ (\bibinfo  {publisher} {Johns Hopkins Univ. Press},\ \bibinfo {year} {1993})\BibitemShut {NoStop}%
\bibitem [{\citenamefont {Cousins}\ \emph {et~al.}(2008)\citenamefont {Cousins}, \citenamefont {Linnemann},\ and\ \citenamefont {Tucker}}]{Cousins:2007bmb}%
  \BibitemOpen
  \bibfield  {author} {\bibinfo {author} {\bibfnamefont {R.~D.}\ \bibnamefont {Cousins}}, \bibinfo {author} {\bibfnamefont {J.~T.}\ \bibnamefont {Linnemann}},\ and\ \bibinfo {author} {\bibfnamefont {J.}~\bibnamefont {Tucker}},\ }\bibfield  {title} {\bibinfo {title} {{Evaluation of three methods for calculating statistical significance when incorporating a systematic uncertainty into a test of the background-only hypothesis for a Poisson process}},\ }\href {https://doi.org/10.1016/j.nima.2008.07.086} {\bibfield  {journal} {\bibinfo  {journal} {Nucl. Instrum. Meth. A}\ }\textbf {\bibinfo {volume} {595}},\ \bibinfo {pages} {480} (\bibinfo {year} {2008})},\ \Eprint {https://arxiv.org/abs/physics/0702156} {arXiv:physics/0702156} \BibitemShut {NoStop}%
\bibitem [{\citenamefont {Neyman}\ and\ \citenamefont {Scott}(1948)}]{Neyman:1948}%
  \BibitemOpen
  \bibfield  {author} {\bibinfo {author} {\bibfnamefont {J.}~\bibnamefont {Neyman}}\ and\ \bibinfo {author} {\bibfnamefont {E.~L.}\ \bibnamefont {Scott}},\ }\bibfield  {title} {\bibinfo {title} {{Consistent Estimates Based on Partially Consistent Observations}},\ }\href {https://doi.org/10.2307/1914288} {\bibfield  {journal} {\bibinfo  {journal} {Econometrica}\ }\textbf {\bibinfo {volume} {16}},\ \bibinfo {pages} {1} (\bibinfo {year} {1948})}\BibitemShut {NoStop}%
\bibitem [{\citenamefont {Murphy}\ and\ \citenamefont {van~der Vaart}(2000)}]{Murphy:2000profile}%
  \BibitemOpen
  \bibfield  {author} {\bibinfo {author} {\bibfnamefont {S.~A.}\ \bibnamefont {Murphy}}\ and\ \bibinfo {author} {\bibfnamefont {A.~W.}\ \bibnamefont {van~der Vaart}},\ }\bibfield  {title} {\bibinfo {title} {{On profile likelihood}},\ }\href {https://doi.org/10.1080/01621459.2000.10474219} {\bibfield  {journal} {\bibinfo  {journal} {J. Amer. Statist. Assoc.}\ }\textbf {\bibinfo {volume} {95}},\ \bibinfo {pages} {449} (\bibinfo {year} {2000})}\BibitemShut {NoStop}%
\bibitem [{\citenamefont {Newey}(1994)}]{Newey:1994}%
  \BibitemOpen
  \bibfield  {author} {\bibinfo {author} {\bibfnamefont {W.~K.}\ \bibnamefont {Newey}},\ }\bibfield  {title} {\bibinfo {title} {{The Asymptotic Variance of Semiparametric Estimators}},\ }\href {https://doi.org/10.2307/2951752} {\bibfield  {journal} {\bibinfo  {journal} {Econometrica}\ }\textbf {\bibinfo {volume} {62}},\ \bibinfo {pages} {1349} (\bibinfo {year} {1994})}\BibitemShut {NoStop}%
\bibitem [{\citenamefont {Chernozhukov}\ \emph {et~al.}(2018)\citenamefont {Chernozhukov}, \citenamefont {Chetverikov}, \citenamefont {Demirer}, \citenamefont {Duflo}, \citenamefont {Hansen}, \citenamefont {Newey},\ and\ \citenamefont {Robins}}]{Chernozhukov:2018dml}%
  \BibitemOpen
  \bibfield  {author} {\bibinfo {author} {\bibfnamefont {V.}~\bibnamefont {Chernozhukov}}, \bibinfo {author} {\bibfnamefont {D.}~\bibnamefont {Chetverikov}}, \bibinfo {author} {\bibfnamefont {M.}~\bibnamefont {Demirer}}, \bibinfo {author} {\bibfnamefont {E.}~\bibnamefont {Duflo}}, \bibinfo {author} {\bibfnamefont {C.}~\bibnamefont {Hansen}}, \bibinfo {author} {\bibfnamefont {W.}~\bibnamefont {Newey}},\ and\ \bibinfo {author} {\bibfnamefont {J.}~\bibnamefont {Robins}},\ }\bibfield  {title} {\bibinfo {title} {{Double/debiased machine learning for treatment and structural parameters}},\ }\href {https://doi.org/10.1111/ectj.12097} {\bibfield  {journal} {\bibinfo  {journal} {Econom. J.}\ }\textbf {\bibinfo {volume} {21}},\ \bibinfo {pages} {C1} (\bibinfo {year} {2018})}\BibitemShut {NoStop}%
\bibitem [{\citenamefont {Dauncey}\ \emph {et~al.}(2014)\citenamefont {Dauncey}, \citenamefont {Sherwood}, \citenamefont {Sherwood},\ and\ \citenamefont {sherwood}}]{Dauncey:2014xga}%
  \BibitemOpen
  \bibfield  {author} {\bibinfo {author} {\bibfnamefont {P.~D.}\ \bibnamefont {Dauncey}}, \bibinfo {author} {\bibfnamefont {M.}~\bibnamefont {Sherwood}}, \bibinfo {author} {\bibfnamefont {S.}~\bibnamefont {Sherwood}},\ and\ \bibinfo {author} {\bibfnamefont {A.}~\bibnamefont {sherwood}},\ }\bibfield  {title} {\bibinfo {title} {{Handling uncertainties in background shapes: the discrete profiling method}},\ }\href@noop {} {\bibfield  {journal} {\bibinfo  {journal} {JINST}\ }\textbf {\bibinfo {volume} {9}},\ \bibinfo {pages} {P04005}},\ \Eprint {https://arxiv.org/abs/1408.6865} {arXiv:1408.6865} \BibitemShut {NoStop}%
\end{thebibliography}%

\appendix

\section{Posterior calibration}
\label{app:calibration}

The GP posterior provides a point estimate and an uncertainty band for the true template rate.  
For these to be useful in a likelihood, the uncertainty must be well-calibrated: the $68\%$ posterior interval should contain the truth approximately $68\%$ of the time across repeated MC realisations.

Table~\ref{tab:calibration} reports the mean per-bin coverage of the GP posterior interval and the histogram $1/\sqrt{a_j}$ interval over $100$ MC realisations at $N_{\rm MC}=2000$ for five truth distributions chosen to probe different shape classes: exponential, flat, steep exponential, Gaussian peak, and quadratic rise.
Both methods recover the nominal $68.3\%$ coverage to within sampling fluctuation across every shape, with the GP achieving comparable coverage to the histogram without inheriting the bin-to-bin fluctuations of $1/\sqrt{a_j}$ that contaminate the histogram band in low-statistics bins. 
The same check at $N_{\rm MC} \in \{200, 500, 1000, 2000, 5000\}$ for the exponential truth gives GP coverages in the range $66\%$--$69\%$ and histogram coverages in the range $67\%$--$69\%$, both consistent with nominal across the
sample-size grid.

\begin{table}[htb!]
\centering
\caption{Mean bin-level $68\%$ coverage across $100$ MC realisations at $N_{\rm MC}=2000$ for five truth distributions.
Both the GP posterior interval and the histogram $1/\sqrt{a_j}$ interval are consistent with the nominal $68.3\%$.}
\label{tab:calibration}
\begin{tabular}{l r r}
\hline\hline
Shape & GP & Histogram \\
\hline
Exponential    & $67.5\%$ & $68.7\%$ \\
Flat           & $67.2\%$ & $67.7\%$ \\
Steep exp.     & $67.8\%$ & $68.2\%$ \\
Gaussian peak  & $68.7\%$ & $68.3\%$ \\
Quadratic rise & $67.4\%$ & $67.4\%$ \\
\hline\hline
\end{tabular}
\end{table}

\section{Eigenmode structure and Barlow--Beeston equivalence}
\label{app:eigenstructure}

\subsection{Eigenvalue spectrum and mode shapes}

\begin{figure}[htb!]
\centering
\includegraphics[width=\columnwidth]{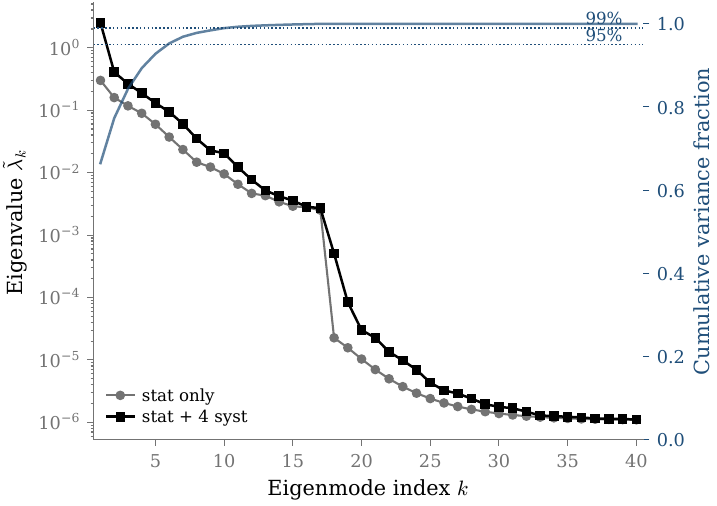}
\caption{Eigenvalue spectrum for Experiment~B\@.  Circles: stat
only.  Squares: stat + four systematics.  Right axis: cumulative
variance fraction.}
\label{fig:eigenspectrum}
\end{figure}

Figure~\ref{fig:eigenspectrum} shows the eigenvalue spectrum of the
GP posterior covariance for Experiment~B, with and without systematics.
In the stat-only case the eigenvalues decay smoothly from
$\lambda_1=0.30$ across the $40$ bins.  Adding the four systematic
directions inserts a dominant calibration mode at
$\tilde\lambda_1=2.50$ followed by three modes of order
$0.2$--$0.4$ that capture mixtures of the resolution and
normalisation systematics.  The cumulative variance fraction
reaches $95\%$ at $k=6$ and $99\%$ at $k=11$ for the combined
covariance.

\begin{figure*}[htb!]
\centering
\includegraphics[width=\textwidth]{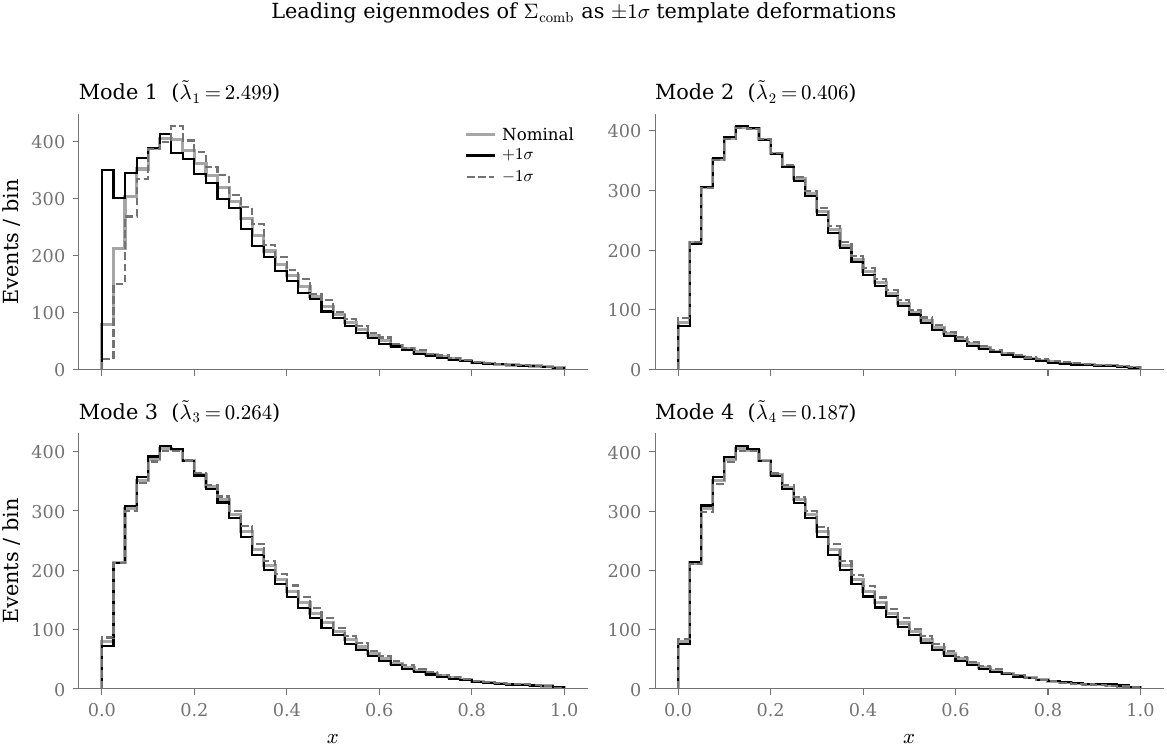}
\caption{Leading four eigenmodes of the combined covariance for
Experiment~B, visualised as $\pm 1\sigma$ template deformations.
Each mode predominantly captures one systematic direction.}
\label{fig:mode_shapes}
\end{figure*}

Figure~\ref{fig:mode_shapes} shows the four leading eigenmodes of the
combined covariance as $\pm 1\sigma$ template deformations.  Each
mode is visually identifiable with its corresponding systematic
direction, confirming that the rank-1 updates of
Eq.~\eqref{eq:combined_cov} are faithfully captured.

\subsection{Mode-truncation behaviour and full-rank equivalence}

\begin{figure}[htb!]
\centering
\includegraphics[width=\columnwidth]{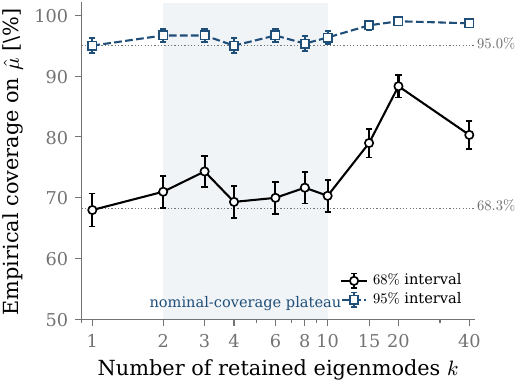}
\caption{Empirical interval coverage of $\hat\mu$ as a function of
the number of retained eigenmodes $k$, on $300$ pseudo-experiments
per $k$ drawn from Dataset~B at $\mu_{\rm true}=1$.  Open circles
are the $68\%$ interval, open squares the $95\%$ interval; error
bars are binomial standard errors.  Dotted horizontal lines mark
the nominal $68.3\%$ and $95.0\%$.  The shaded vertical band is
the plateau $k\in[2,10]$ on which both intervals deliver nominal
coverage; under-truncation below $k=2$ drops variance and
undercovers, while over-truncation toward the $44$-parameter
histogram budget inflates $\sigma_{\hat\mu}$ and overcovers.}
\label{fig:cov_vs_k}
\end{figure}

The analytical argument of Section~\ref{sec:bb_equiv} predicts that
full-rank GP eigenmodes reproduce Barlow--Beeston inference.
Figure~\ref{fig:cov_vs_k} maps the truncation-versus-coverage
tradeoff on Dataset~B, fitting the eigenmode model at
$k\in\{1,2,3,4,6,8,10,15,20,40\}$ on $300$ pseudo-experiments per $k$
at $\mu_{\rm true}=1$.  Inside the plateau $k\in[2,10]$ both the
$68\%$ and $95\%$ interval coverages sit at the nominal $68.3\%$
and $95.0\%$ within sampling error.  Below $k=2$ the truncation
drops genuine variance and the intervals undercover.  Above
$k\approx 12$ the fit retains statistical modes that the data do
not constrain and the parabolic Hesse error widens, recovering the
overcoverage of the full $44$-parameter BB+histosys benchmark in
Section~\ref{sec:exp_b}.  The intermediate plateau is the operating
range that the $95\%$-variance threshold automatically selects; its
breadth ($k$ between $2$ and $10$ all deliver nominal coverage)
shows that the threshold does not need fine tuning.

\section{HistFactory implementation agreement}
\label{app:histfactory}

We compare the GP eigenmode parameterisation against the full
histogram-based HistFactory modifier chain on a 15-bin single-channel
benchmark with a histosys (slope) shape systematic, a 10\% normsys,
and per-bin staterror.  The HistFactory model carries
$15+2+1=18$ parameters; the GP eigenmode model carries $k+2+1$,
with $k$ chosen by the 99\% variance threshold.  Two thousand
pseudo-experiments are generated with a fixed shape shift
($\alpha_{\rm true}=0.3$) at $\mu_{\rm true}=1$.

\begin{figure}[htb!]
\centering
\includegraphics[width=\columnwidth]{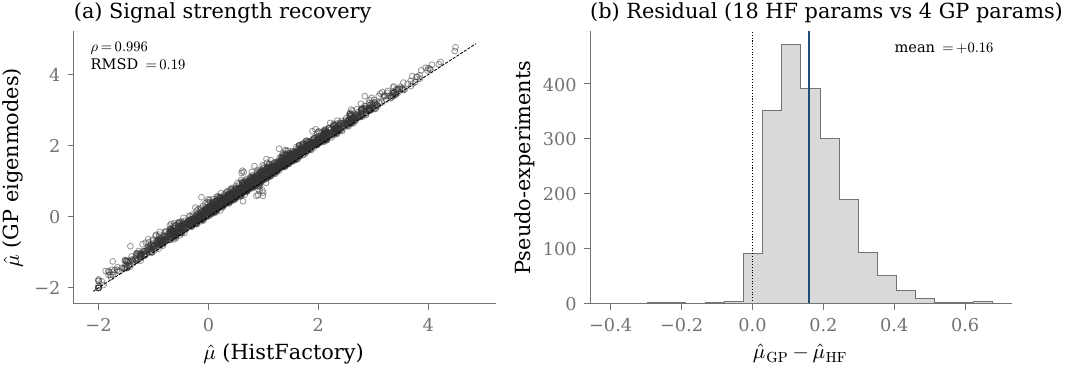}
\caption{Modifier chain recovery: (a)~$\hat\mu_{\rm GP}$ versus
  $\hat\mu_{\rm HF}$ for $2000$ pseudo-experiments with shape shift
  $\alpha_{\rm true}=0.3$, correlation $\rho=0.996$, root-mean-square
  residual $0.19$.  (b)~Histogram of $\hat\mu_{\rm GP}-\hat\mu_{\rm HF}$
  with empirical mean $+0.16$, the plug-in penalty quantified in
  Appendix~\ref{app:semiparametric}.}
\label{fig:modifier_chain}
\end{figure}

\begin{figure}[htb!]
\centering
\includegraphics[width=\columnwidth]{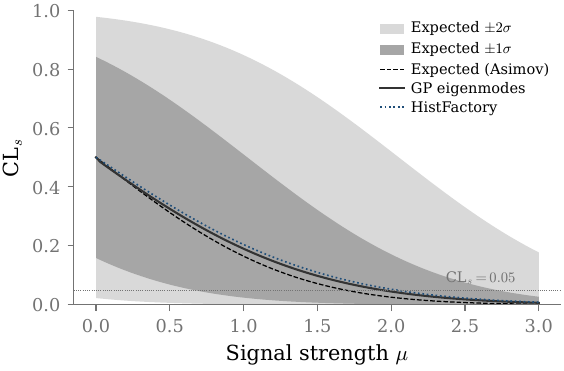}
\caption{Asymptotic CL$_s$ scan on the same benchmark.  GP
  eigenmodes (circles) and HistFactory (squares) overlap inside the
  Asimov $\pm 1\sigma$ band; the 95\% upper limits agree at the
  percent level.}
\label{fig:cls}
\end{figure}

Figure~\ref{fig:modifier_chain}~(a) plots $\hat\mu_{\rm GP}$ against
$\hat\mu_{\rm HF}$ for each pseudo-experiment.  The correlation is
$\rho=0.996$, confirming that the eigenmode parameterisation tracks
the modifier chain pseudo-experiment by pseudo-experiment.  The
root-mean-square residual on $\hat\mu_{\rm GP}-\hat\mu_{\rm HF}$ is
$0.19$ with a residual mean of $+0.16$.  The non-zero mean is the
plug-in penalty disclosed in Appendix~\ref{app:semiparametric}: the
GP smoothes the nominal background once and propagates the smoothed
shape to all pseudo-experiments, while HistFactory profiles the slope
nuisance parameter against each pseudo-experiment's data.  The two
methods are equivalent in the high-MC, no-shape-shift limit
(Sec.~\ref{sec:bb_equiv}, Eq.~\eqref{eq:variance_bound}); the
residual reported here is the cost of the two-step plug-in at the
shape-shift magnitude tested.  Figure~\ref{fig:cls} shows the
corresponding CL$_s$ scan: the GP and HistFactory observed lines
follow each other within the $\pm 1\sigma$ Asimov band, and the 95\%
upper limits agree at the percent level.

\section{Semiparametric efficiency and the profiling gap}
\label{app:semiparametric}

The GP eigenmode method produces templates with lower variance than
Barlow--Beeston (Eq.~\eqref{eq:variance_bound}), uses fewer nuisance
parameters (Sec.~\ref{sec:exp_b}), and eliminates spurious
signals from MC noise.  These properties do not, however,
automatically translate into higher discovery significance per
measured event.  The reason is structural: the GP template is a
two-step plug-in estimator, while Barlow--Beeston performs a joint
profile that achieves the semiparametric efficiency bound.  This
appendix derives the bound, shows that BB attains it, identifies the
variance penalty incurred by the GP plug-in, and clarifies the role of
the DML correction in this context.

\subsection{The Poisson semiparametric model}

A binned signal-plus-background analysis is a semiparametric inference
problem in the sense of Bickel, Klaassen, Ritov, and
Wellner~\cite{Bickel:1993}.  The parameter of interest is $\mu$.  The
nuisance is the background shape $\boldsymbol{g} = (g_1, \ldots,
g_N)$, an $N$-vector whose true value is unknown.  The observed data
and the MC auxiliary measurement are
\begin{align}
n_j &\sim \Poisson(\mu\, s_j + g_j)\,,
\label{eq:app_data} \\
a_j &\sim \Poisson(\tau\, g_j)\,,
\label{eq:app_mc}
\end{align}
where $\tau$ is the MC-to-data luminosity ratio and $s_j$ is the
known signal template.  All $2N$ observations are mutually
independent.  The full log-likelihood is
\begin{equation}
\ell(\mu, \boldsymbol{g})
= \sum_{j=1}^{N} \bigl[
  n_j \ln(\mu s_j + g_j) - (\mu s_j + g_j)
  + a_j \ln(\tau g_j) - \tau g_j
\bigr]\,.
\label{eq:app_full_ll}
\end{equation}
This is the multi-bin generalisation of the ``on/off'' problem treated
by Cousins, Linnemann, and Tucker~\cite{Cousins:2007bmb}.  Because the
number of nuisance parameters grows with the number of bins, the
problem lies in the incidental-parameters regime of Neyman and
Scott~\cite{Neyman:1948}, and a na\"{\i}ve treatment of the $g_j$ as
fixed unknowns would produce inconsistent inference.  The constraint
provided by the auxiliary measurement~\eqref{eq:app_mc} regularises
the problem.

\subsection{The semiparametric information bound}

The Fisher information matrix of the model
\eqref{eq:app_data}--\eqref{eq:app_mc} at the true parameter values
$(\mu_0, \boldsymbol{g}_0)$ has block structure
\begin{equation}
\mathcal{I} = \begin{pmatrix}
I_{\mu\mu} & I_{\mu g}^T \\[2pt]
I_{\mu g}  & I_{gg}
\end{pmatrix}\,,
\label{eq:app_fisher}
\end{equation}
with entries
\begin{align}
I_{\mu\mu} &= \sum_j \frac{s_j^2}{\nu_j}\,,
\qquad
[I_{\mu g}]_j = \frac{s_j}{\nu_j}\,,
\nonumber \\
[I_{gg}]_{jk} &= \delta_{jk}\!\left(
  \frac{1}{\nu_j} + \frac{\tau}{g_j}\right)\,,
\label{eq:app_fisher_entries}
\end{align}
where $\nu_j = \mu_0 s_j + g_j$.  The diagonal structure of $I_{gg}$
reflects the independence of Poisson bins.  Eliminating
$\boldsymbol{g}$ by the Schur complement gives the efficient Fisher
information for $\mu$:
\begin{equation}
I_{\mu\mu}^{\rm eff}
= I_{\mu\mu} - I_{\mu g}^T\, I_{gg}^{-1}\, I_{\mu g}\,.
\label{eq:app_schur}
\end{equation}
Because $I_{gg}$ is diagonal, the matrix inversion and multiplication
reduce to a single sum.  Writing $[I_{gg}^{-1}]_{jj} = \nu_j\, g_j
/ (\tau\nu_j + g_j)$ and substituting into~\eqref{eq:app_schur}
yields, after simplification,
\begin{equation}
I_{\mu\mu}^{\rm eff}
= \sum_j \frac{s_j^2}{\nu_j}
  \;\frac{\tau\,\nu_j / g_j}
       {1 + \tau\,\nu_j / g_j}.
\label{eq:app_Ieff}
\end{equation}

The factor $r_j / (1 + r_j)$ with $r_j = \tau\nu_j/g_j$ is strictly
less than unity for any finite $\tau$, so
$I_{\mu\mu}^{\rm eff} < I_{\mu\mu}^{\rm known} = \sum s_j^2/\nu_j$.
The efficient information for $\mu$ is reduced relative to what would
be available if the background were known exactly.  As $\tau \to
\infty$ the factor approaches unity and the bound converges to the
Cram\'er--Rao bound of the parametric model with known background.

Under the null hypothesis $\mu_0 = 0$ and $\nu_j = g_j$, the factor
simplifies to $\tau/(1+\tau)$ uniformly across bins, giving
\begin{equation}
I_{\mu\mu}^{\rm eff}\big|_{\mu=0}
= \frac{\tau}{1+\tau}\;\sum_j \frac{s_j^2}{g_j}\,.
\label{eq:app_Ieff_null}
\end{equation}
The ratio of efficient to known-background information is therefore
$\tau/(1+\tau)$: for $\tau = 1$ only half the information survives,
for $\tau = 2.5$ (Dataset~A) roughly 71\%, for $\tau = 10$
(Dataset~B) roughly 91\%.  The semiparametric variance bound on any
regular estimator of $\mu$ is
$(I_{\mu\mu}^{\rm eff})^{-1}$.

\subsection{Barlow--Beeston achieves the bound}

The profile likelihood estimator in the full model of
Eq.~\eqref{eq:app_full_ll} maximises $\ell(\mu, \boldsymbol{g})$
jointly over $\mu$ and $\boldsymbol{g}$.  At each fixed $\mu$, the
first-order condition $\partial\ell/\partial g_j = 0$ yields the
profiled background
\begin{equation}
\begin{aligned}
B_j &\equiv n_j + a_j - (1{+}\tau)\,\mu s_j\,,\\
\hat{g}_j(\mu)
&= \frac{B_j + \sqrt{B_j^2 + 4(1{+}\tau)\,\tau\mu s_j a_j}}
        {2(1{+}\tau)}\,,
\end{aligned}
\label{eq:app_ghat}
\end{equation}
which is the closed-form expression of
Conway~\cite{Conway:2011in}.  Substituting back into the
log-likelihood gives the profile likelihood
$\ell_P(\mu) = \ell(\mu, \hat{\boldsymbol{g}}(\mu))$.

Murphy and van~der~Vaart~\cite{Murphy:2000profile} established that,
under regularity conditions satisfied by the Poisson model, the
profile likelihood ratio statistic is asymptotically $\chi^2_1$ and
the resulting MLE achieves the semiparametric bound
$(I_{\mu\mu}^{\rm eff})^{-1}$.  The per-bin gamma factors of
Barlow--Beeston are the reparameterisation $\gamma_j =
\hat{g}_j/g_j^{\rm nom}$ of this joint profiling.

The mechanism by which BB achieves efficiency is worth stating
explicitly.  When a positive fluctuation occurs in a bin overlapping
the signal region, the profiled $\hat{g}_j$ shifts downward,
constrained by the auxiliary $a_j$, to partially absorb the
fluctuation.  This shift improves the background estimate precisely
where it matters for signal extraction.  The joint optimisation
allocates information between background estimation and signal
estimation in the way that minimises the total variance of
$\hat\mu$, which is why it saturates the bound.

\subsection{The GP plug-in as a two-step estimator}

The GP template is constructed in two steps.  First, a GP is fitted to
the MC counts $\boldsymbol{a}$, yielding a smooth background estimate
$\hat{g}_j^{\GP} = e^{\hat{f}(x_j)} w_j / \tau$.  Second, this
estimate is plugged into the data likelihood and $\mu$ is estimated
from
\begin{equation}
\hat\mu^{\rm plug} = \arg\max_\mu
  \sum_j \bigl[
    n_j \ln(\mu s_j + \hat{g}_j^{\GP})
    - (\mu s_j + \hat{g}_j^{\GP})
  \bigr]\,.
\label{eq:app_plugin}
\end{equation}
The background $\hat{g}_j^{\GP}$ does not depend on $\boldsymbol{n}$
and is not adjusted during the maximisation over $\mu$.  The eigenmode
constraint of Eq.~\eqref{eq:gp_expected} partially relaxes this
rigidity by allowing the template to shift along the pre-computed
eigenvectors $\{v_i\}$, but these directions are determined by the GP
posterior covariance, a property of the kernel and the MC sample
$\boldsymbol{a}$, and not by the data-template residuals.

Newey~\cite{Newey:1994} proved a general result for this class of
estimators.  If $\hat\mu^{\rm plug}$ is obtained by plugging a
first-step nonparametric estimate $\hat\eta$ of the nuisance into the
score equation for $\mu$, then $\hat\mu^{\rm plug}$ is
$\sqrt{n}$-consistent with asymptotic variance
$(I_{\mu\mu}^{\rm eff})^{-1} + \Delta^2$, where the non-negative term
$\Delta^2$ vanishes if and only if the plug-in satisfies a
Neyman-orthogonality condition with respect to the nuisance.  For a
generic smooth plug-in, including the GP posterior mean, $\Delta^2 > 0$.

The concrete mechanism is visible in the score function.  The plug-in
score at the true $\mu_0$ is
\begin{equation}
S_\mu^{\rm plug}
= \sum_j s_j\!\left(
  \frac{n_j}{\mu_0 s_j + \hat{g}_j^{\GP}} - 1\right)\,.
\label{eq:app_plug_score}
\end{equation}
Under a perturbation $\hat{g}_j^{\GP} = g_j + \delta g_j$ the
expectation of this score acquires a bias proportional to $-\sum
s_j\,\delta g_j/\nu_j$.  This is nonzero whenever the template
error has any projection onto the ``signal direction''
$s_j/\nu_j$ in bin space.  The GP smoothing of a falling spectrum
generically produces such a projection: smoothing raises the
estimated rate in some bins and lowers it in others, and unless the
signal shape happens to be orthogonal to this perturbation pattern,
$\hat\mu$ is biased.  The sign is systematic: smoothing a concave
spectrum into the signal region biases $\hat\mu$ downward.

The profiled score, by contrast, contains an additional term from the
implicit dependence of $\hat{g}_j$ on $\mu$:
\begin{equation}
S_\mu^{\rm BB}
= \sum_j s_j\!\left(
  \frac{n_j}{\mu s_j + \hat{g}_j(\mu)} - 1\right)
+ \sum_j \frac{s_j}{\nu_j}\,
  \frac{\partial\hat{g}_j}{\partial\mu}\,.
\label{eq:app_bb_score}
\end{equation}
The second term is the Neyman-orthogonal correction: it cancels the
first-order bias from template error.  This is why the profile
likelihood achieves the efficient bound without requiring any explicit
debiasing step.

\subsection{DML: unbiasedness without efficiency}

The DML framework of Chernozhukov \textit{et
al.}~\cite{Chernozhukov:2018dml} constructs an explicit
Neyman-orthogonal moment condition by adding a correction that
exploits the unbiased histogram estimator $\tilde{a}_j = a_j/\tau$:
\begin{equation}
\psi^{\rm orth}
= \sum_j s_j\!\left(\frac{n_j}{\nu_j} - 1\right)
+ \sum_j \frac{s_j}{\nu_j}
  \bigl(\hat{g}_j^{\GP} - \tilde{a}_j\bigr)\,.
\label{eq:app_dml}
\end{equation}
Under a perturbation $\hat{g}^{\GP} = g + \delta g$, the bias from
the first sum and the correction from the second cancel at first
order, as shown in the score-bias derivation above.
The resulting estimator
$\hat\mu^{\rm DML}$ is unbiased and $\sqrt{n}$-consistent.

Its asymptotic variance, however, exceeds
$(I_{\mu\mu}^{\rm eff})^{-1}$.  The correction term introduces
the noise of $\tilde{a}_j$, which has variance $g_j/\tau$ per bin.
A calculation along the lines of Newey~\cite{Newey:1994} gives
\begin{equation}
\Var(\hat\mu^{\rm DML})
= (I_{\mu\mu}^{\rm eff})^{-1}
+ \frac{\sum_j s_j^2\, g_j / (\tau\,\nu_j^2)}
       {\bigl(\sum_k s_k^2/\nu_k\bigr)^2}\,.
\label{eq:app_dml_var}
\end{equation}
The second term is strictly positive for any finite $\tau$.  DML
removes the bias from the plug-in procedure but does not recover the
information that the profile likelihood extracts by jointly adapting
the background to the data.  The net effect is an estimator that is
more robust than the uncorrected plug-in (its coverage is correct) but
less powerful than BB (its confidence intervals are wider).

This result has a clean interpretation in the language of
semiparametric efficiency~\cite{Bickel:1993}.  The efficient estimator
of $\mu$ uses the efficient score, which is the projection of the
full score onto the orthogonal complement of the nuisance tangent
space.  The profile likelihood constructs this projection implicitly
through joint optimisation.  DML constructs it explicitly through the
correction term, but the correction introduces extraneous noise from
the unbiased-but-noisy estimator $\tilde{a}_j$, inflating the
variance beyond the bound.

\subsection{Connection to discrete profiling}

Dauncey \textit{et al.}~\cite{Dauncey:2014xga} addressed a related
problem: uncertainty in the functional form of a smooth background.
Their discrete profiling method treats the function-form index as a
discrete nuisance parameter, profiled alongside $\mu$ by taking
the likelihood envelope across candidate functions.  The coverage
and bias of this method are shown to be good across a range of
scenarios.

Discrete profiling and Barlow--Beeston embody the same structural
principle.  BB profiles over piecewise-constant background shapes,
with one parameter per bin constrained by the MC auxiliary
measurement.  Discrete profiling profiles over parametric shapes,
with the function index penalised by an information criterion.
Both methods allow the background to adapt to the data during
signal extraction.  The GP template, by contrast, commits to a
single smooth shape at the first step and does not revisit that
commitment.  In the language of the semiparametric
literature~\cite{Bickel:1993}, BB and discrete profiling both
estimate the nuisance parameter adaptively within the likelihood,
while the GP plug-in estimates it in a separate preliminary stage.

\subsection{Quantifying the significance deficit}

\begin{figure}[htb!]
\centering
\includegraphics[width=\columnwidth]{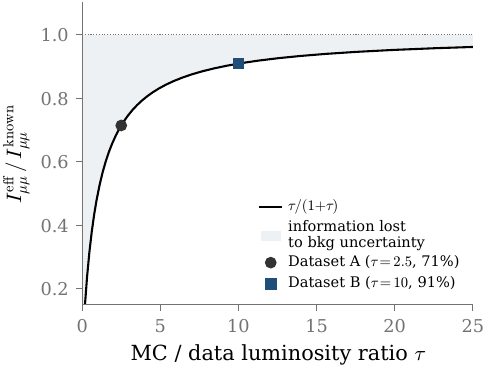}
\caption{The efficient information fraction
$I_{\mu\mu}^{\rm eff}/I_{\mu\mu}^{\rm known} = \tau/(1+\tau)$ as a
function of the MC-to-data luminosity ratio.  The shaded region is
the information consumed by background uncertainty.  Dataset~A
operates at $\tau=2.5$, where $29\%$ of the signal-strength
information is lost to background uncertainty; Dataset~B at
$\tau=10$, where the loss is $9\%$.  The curve is the upper bound
on the variance penalty any plug-in estimator can pay relative to
the joint profile.}
\label{fig:info_frac}
\end{figure}

\begin{figure}[htb!]
\centering
\includegraphics[width=\columnwidth]{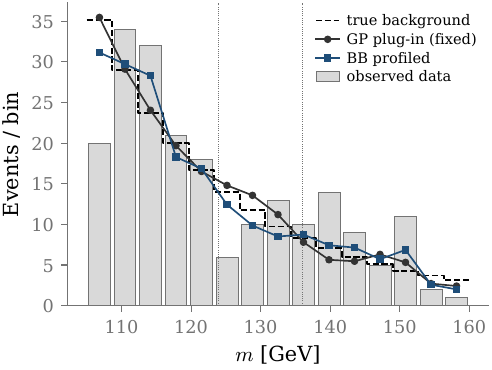}
\caption{Background estimates for a single Dataset~A pseudo-experiment
at $\mu_{\rm true}=1$.  The signal region $\mu_s\pm 3\sigma_s$ is
marked by dotted vertical lines.  The GP plug-in background
(circles) is fixed at its pre-fit shape and does not respond to the
observed counts in the signal region.  The Barlow--Beeston profiled
background (squares) shifts downward inside the signal region,
absorbing part of the data--template discrepancy jointly with $\mu$
through Conway's closed-form solution Eq.~\eqref{eq:app_ghat}.  This
local adaptation is the mechanism by which the joint profile
attains the semiparametric bound.}
\label{fig:bkg_adaptation}
\end{figure}

The semiparametric analysis is more than a structural caveat: it
decides which method is appropriate at which $\tau$, and the
quantitative answers come from two complementary pictures.  The
first quantifies how much of the parameter-of-interest information
remains after the nuisance has been profiled (Fig.~\ref{fig:info_frac}).
The second exhibits, on a single pseudo-experiment, the mechanism by
which the BB joint profile recovers that information while the GP
plug-in cannot (Fig.~\ref{fig:bkg_adaptation}).  Together they
justify the regime split asserted in the main-text conclusions: BB
is structurally the right tool in the single-channel statistically
limited regime; the GP eigenmode method dominates in the
systematically-limited and high-dimensional regimes where the
efficiency loss is bounded above by $1-\tau/(1+\tau)$, which is
small at the $\tau$ values typical of those analyses.

The efficiency ratio between the GP plug-in and BB can be expressed
in terms of the template error projection.  Under the null, the BB
significance scales as
\begin{equation}
Z_{\rm BB}
\approx Z_{\rm known}\;\sqrt{\frac{\tau}{1+\tau}}\,,
\label{eq:app_Zbb}
\end{equation}
where $Z_{\rm known}$ is the significance with perfectly known
background.  The GP plug-in has access to the full
known-background information $\sum s_j^2/\nu_j$ but evaluates it at
the wrong background, producing
\begin{equation}
Z_{\GP}^{\rm plug}
\approx Z_{\rm known}\;\sqrt{\frac{\tau}{1+\tau}}
\;\left(1
  - \frac{\sum_j s_j\,\delta g_j/\nu_j}{\sum_j s_j^2/\nu_j}
\right)\,,
\label{eq:app_Zgp}
\end{equation}
where $\delta g_j = \hat{g}_j^{\GP} - g_j$ is the template error.
The fractional significance deficit is therefore
\begin{equation}
\frac{Z_{\rm BB} - Z_{\GP}}{Z_{\rm BB}}
\;\approx\;
\frac{\langle\delta\boldsymbol{g},\,
      \boldsymbol{s}/\boldsymbol{\nu}\rangle}
     {\|\boldsymbol{s}/\boldsymbol{\nu}\|^2}\,,
\label{eq:app_frac_loss}
\end{equation}
which is the cosine-weighted projection of the template error onto
the signal direction in the inner product
$\langle u, v\rangle = \sum u_j v_j$.  For a smoothed falling
spectrum with a localised signal, this projection is generically
of order $\ell/L$ where $\ell$ is the GP lengthscale and $L$ is
the domain length.  A lengthscale of 10~GeV in a 55~GeV domain
gives a fractional loss of order 0.18, consistent with the
$\sim 0.3\sigma$ deficit observed in the benchmark of
Sec.~\ref{sec:exp_a} at $Z \approx 1.5\sigma$.

The eigenmode constraint reduces this projection by absorbing
template error along the leading eigenvectors.  The residual
projection depends on the alignment between the signal direction
$s_j/\nu_j$ and the span of the retained eigenmodes.  For a narrow
signal in a smoothly falling spectrum, the signal direction is a
localised bump while the eigenmodes are smooth, extended
oscillations, so the alignment is generically poor and the
residual projection remains nonzero.

\subsection{Regime summary}

The analysis above delineates where each method has the advantage.
In the statistically limited, single-channel regime, BB's joint
profiling achieves the semiparametric efficiency bound and the GP
plug-in does not; the per-bin gamma factors are not wasted
parameters but the mechanism of efficient inference.  In the
systematically limited regime, the efficiency loss from template
uncertainty is negligible ($\tau/(1+\tau) \approx 1$ for large $\tau$)
and the dominant concern is the parameterisation of shape variations,
where eigenmode compression provides genuine gains in parameter count
and interpretability.  In high-dimensional template spaces, BB
requires one gamma per bin with the count growing exponentially in the
number of observables, while the GP eigenmode count grows with the
effective dimensionality of the kernel; for a three-dimensional
template with $20^3$ bins, BB introduces 8000 nuisance parameters
while the GP method might need 20--50.  Finally, the smooth GP
background provides the covariance structure of the test-statistic
field analytically (via the GP kernel and the Rice formula
for upcrossings, see~\cite{Gross:2010qma,Gandrakota:2022gplee}),
enabling computation of the look-elsewhere trials factor without
Monte Carlo, a capability that BB does not offer.

\end{document}